# *J*-driven Dynamic Nuclear Polarization for sensitizing high field solution state NMR

Maria Grazia Concilio[1*], Ilya Kuprov[2], Lucio Frydman,[1,3*]

[1]Department of Chemical and Biological Physics, Weizmann Institute of Science, Rehovot, Israel.
[2]School of Chemistry, University of Southampton, Southampton, UK.
[3]National High Magnetic Field Laboratory, Tallahassee, Florida, USA.

**Abstract**

Dynamic nuclear polarization (DNP) is widely used to enhance solid state nuclear magnetic resonance (NMR) sensitivity. Its efficiency as a generic signal-enhancing approach for liquid state NMR, however, decays rapidly with magnetic field $B_0$, unless mediated by scalar interactions arising only in exceptional cases. This has prevented a more widespread use of DNP in structural and dynamical solution NMR analyses. This study introduces a potential solution to this problem, relying on biradicals with exchange couplings $J_{ex}$ of the order of the electron Larmor frequency $\omega_E$. Numerical and analytical calculations show that in such $J_{ex} \approx \pm\omega_E$ cases a phenomenon akin to that occurring in chemically induced DNP (CIDNP) happens, leading to different relaxation rates for the biradical singlet and triplet states which are hyperfine-coupled to the nuclear $\alpha$ or $\beta$ states. Microwave irradiation can then generate a transient nuclear polarization build-up with high efficiency, at all magnetic fields that are relevant in contemporary NMR, and for all rotational diffusion correlation times that occur in small- and medium-sized molecules in conventional solvents.

Corresponding authors:

*maria-grazia.concilio@weizmann.ac.il
*lucio.frydman@weizmann.ac.il

## 1. Introduction

Higher NMR sensitivity could bring transformative breakthroughs to analytical, pharmaceutical and biophysical chemistry. NMR sensitivity can be enhanced by higher external magnetic fields $B_0$, but this is a slow, expensive approach. An alternative arises if electron magnetization is transferred, from a stable radical, to the nuclei to be detected. By irradiating electrons with microwaves a their Larmor frequency, $\omega_E$, the so-called Dynamic Nuclear Polarization (DNP) effect can then enhance NMR sensitivity up to the ratio between the gyromagnetic constants of the electron and the nucleus: $\gamma_E/\gamma_N$. Based on the irradiation of a stable organic monoradical, such effect can enhance the Boltzmann equilibrium nuclear magnetization by factors of hundredfold, thereby transforming the analytical potential of NMR. Predicted by Overhauser in 1953[1] and thereafter confirmed by Carver and Slichter,[2] DNP has revolutionized solid state NMR;[3-6] it has also made inroads into *in vivo* spectroscopy, based on rapid melting approaches.[7,8] DNP, however, has not yet impacted what is arguably the widest of NMR realms – high-field solution-state studies. This longstanding problem arises from an Overhauser DNP efficiency, that in liquids depends on the electron-nuclear cross-relaxation rate $\sigma_{E,N}$:

$$\sigma_{E,N} \approx \frac{\gamma_E^2 \gamma_N^2 \hbar^2}{10} \left(\frac{\mu_0}{4\pi}\right)^2 \frac{\tau_C}{r_{EN}^6} \left(\frac{6}{1+(\omega_E+\omega_N)^2 \tau_C^2} - \frac{1}{1+(\omega_E-\omega_N)^2 \tau_C^2}\right). \quad (1)$$

The typical rotational correlation time $\tau_c$ of a radical/nucleus dipolar-coupled spin pair is $\tau_c \approx 0.1\text{-}1$ ns, while typical electron and nuclear Larmor frequencies in mid- to high-field scenarios are $\omega_E \geq 200$ GHz and $\omega_N \geq 300$ MHz. This leads to $(\omega_E \pm \omega_N)^2 \tau_C^2 \gg 1$; i.e, negligible cross-relaxation rates in Eq. (1) for most cases of practical analytical interest, and Overhauser DNP efficiencies that decrease quadratically with $B_0$. Consequently – and unless aided by the contact couplings that can arise for certain radicals and solutes[9-13] – typical $^1$H DNP enhancements drop from a maximum of $\approx 330$x when $B_0 \leq 0.4$T, to $\approx 1.001$x at the $\geq 7$ T fields were contemporary NMR is done.[11,14-17] We recently discussed a way to bypass this bottleneck, proposing a cross-correlated (CC) DNP strategy involving biradicals as polarization sources.[18] CCDNP, however, required significant coincidences between nuclear/electron spin interaction tensors, and long-term stability in the nuclear-electron geometry; for optimal conditions, it then provided steady-state NMR enhancements that approached $\approx 10\text{-}20$x. This study reports a further investigation into the physics of a three-spin biradical/nuclear system, revealing a new polarization transfer possibility. Unlike CCDNP, the mechanism that is here introduced: (i) does not put stringent conditions on multiple independent coupling tensors; (ii) can lead to nuclear polarization enhancements of $\approx 100\text{-}200$x for $B_0$ ranging from $\leq 1$T to $\geq 20$ T and for rotational correlation times in the 0.1 - 1 ns range; and (iii) does not result from a steady state arising upon electron saturation, but rather from transient phenomena. At the centre of this proposal are two electrons interacting through an exchange coupling $J_{ex}$ in the order of $\omega_E$, a chemically tuneable condition that can be fulfilled by many biradicals known to have exchange couplings in the $0 \leq J_{ex} \leq 1$ THz range.[19-21] Under such conditions we found that moderate microwave fields can lead to the DNP enhancement improvements shown in Figure 1. The present study demonstrates and explains the basis of this J-driven (J-DNP) mechanism based on Liouville space numerical simulations and analytical calculations using Redfield's relaxation theory [22-26]. The latter serve to highlight similarities between the roles that the electronic singlet and triplet states play in polarizing nuclei in J-DNP, and those played by singlet and triplet states arising in chemically-induced DNP (CIDNP) experiments.[27-30]

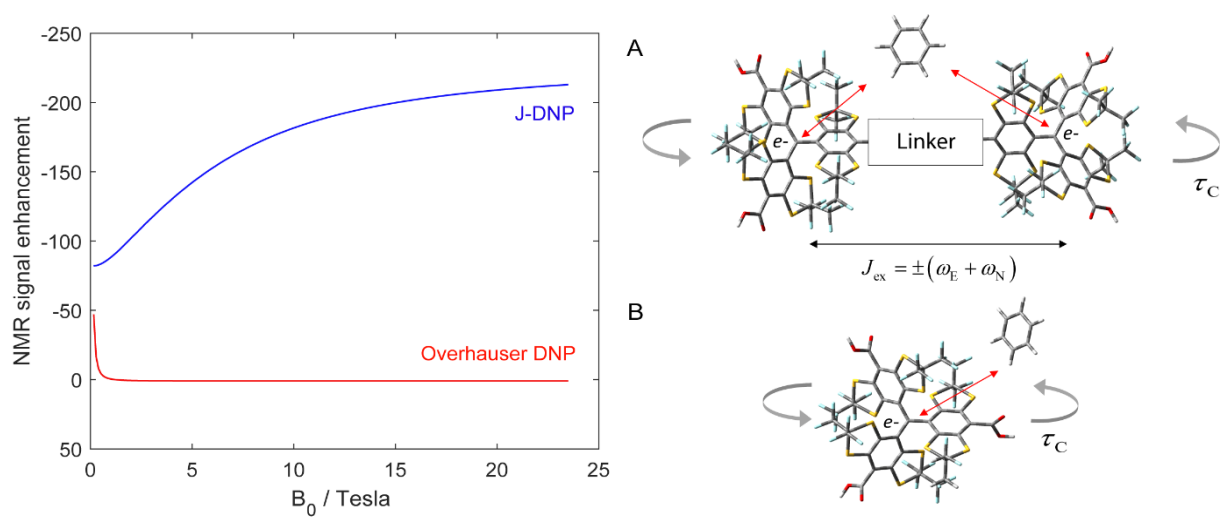

**Fig. 1:** *On the left:* Comparison between the maximal enhancements delivered by the Overhauser and by J-driven DNP, as function of $B_0$. *On the right:* Models of the biradical/nuclear system assumed for simulating J-DNP (A) and of the monoradical/nuclear system used for Overhauser DNP (B). The only nucleus in the system (a proton) was assumed in the "solvent" molecule; the red arrows represent dipolar interactions between the electron(s) placed in the centre of the radical and the proton in this "solvent". No other protons/nuclei were assumed. The J-DNP simulation was performed using a biradical/proton dipolar-coupled triad (A) with a $\tau_c$ = 500 ps rotational correlation time, a $J_{ex}$ = +($\omega_E + \omega_N$) at each field strength, and other parameters as given in Table 1. The Overhauser DNP simulation was performed for a monoradical/proton dipolar-coupled pair (B) with $\tau_c$ = 157 ps (typical of trityl[31]) and other parameter as given in Table 1 –but with one of the electrons in the Table absent. The enhancement in the J-DNP was calculated after 20 ms of microwave irradiation; the Overhauser DNP enhancements were calculated vs field at the steady state. Overhauser enhancements slightly larger than those predicted in the Fig. 1 were observed in water solutions at 1.4 T and 3.4 T;[32,33] this likely arises due to translational diffusion effects, which were not consider in this work.

## 2. Spin system and theoretical methodology

The system examined in this work was a biradical interacting with a proton exclusively through dipolar (*aka* anisotropic hyperfine) couplings. Hyperfine couplings were assumed between both electrons and the proton, the inter-particle distances were assumed fixed. These distances between the two unpaired spin-1/2 electrons (belonging to the radical) and the spin-1/2 proton (belonging to the solvent) were 8.6 Å and 11.1 Å respectively (see Table 1 for the actual proton and electron Cartesian coordinates). Scalar electron-nuclear couplings were set to zero. The J-DNP mechanism described in this study was assumed driven by a rotational dynamics, modulating the relaxation of the three-spin system with a rotational correlation time, $\tau_c$, as shown in Fig. 1. The two electrons are assumed to have identical *g*-tensors, assumed anisotropic for the sake of realism (even if the Supporting Information 6 shows that neither nuclear shielding anisotropy nor *g*-anisotropy are required for the J-DNP enhancement). The most important variable in the system is the isotropic inter-electron exchange coupling $J_{ex}$, which was modelled on the basis of trityl-based symmetric biradicals for which such couplings have been observed.[19,20] These parameters were used to create the spin Hamiltonian, and to calculate the evolution subject to microwave irradiation and relaxation as per Redfield's theory[22,25] (see Supporting Information 1 for additional discussion of these radicals, including their estimated solution-state $T_1$s and $T_2$s[34,35]).

**Table 1:** Biradical / proton parameters used in the simulations shown in Figures 1 - 5. $B_0$, $J_{ex}$ and $\tau_c$ were set as described in the figures; all other coupling parameters relied on the inter-spin distances.

| Parameter | Value |
|---|---|
| $^1$H chemical shift tensor, ppm | [5 10 20] |
| g-tensor[1] for electrons 1 and 2, Bohr magneton | [2.0032 2.0032 2.0026] |
| $^1$H coordinates, [x y z], Å | [-3 0.5 1.3] |
| Electron 1 coordinates, [x y z], Å | [0 0 -9.37] |
| Electron 2 coordinates, [x y z], Å | [0 0 9.37] |
| Scalar relaxation modulation depth, GHz | 3 |
| Scalar relaxation modulation time, ps | 1 |
| Microwave nutation frequency, $\omega_{\mu w}$, MHz | 1 |
| Temperature, K | 298 |

[1] Simulations used identical Zeeman interaction tensors, that were made axially symmetric along the main molecular axis (corresponding the linker connecting the two trityls).

The bulk of this study focuses on a $J_{ex} \gg \omega_{\Delta e}$ scenario, where $\omega_{\Delta e} = \omega_{e1} - \omega_{e2}$ is the difference between the Larmor frequency of the two electrons. In such cases the electron Zeeman eigenstates $|\alpha_{e1}\beta_{e2}>$ and $|\beta_{e1}\alpha_{e2}>$ are no longer eigenfunctions of the spin Hamiltonian; the relevant Hamiltonian was therefore treated in the singlet/triplet electron basis set $\left\{\hat{S}_0^{(e1,e2)}, \hat{T}_0^{(e1,e2)}, \hat{T}_{\pm1}^{(e1,e2)}\right\}$. As shown in Supporting Information 2, this leads to a microwave rotating frame Hamiltonian:

$$\hat{H}_{rot} = \hat{H}^{S_0T_0} + \hat{H}^{T_{+1}T_{-1}} + \hat{H}_{\mu w} \qquad (2)$$

where $\hat{H}^{S_0T_0}$ is a Hamiltonian acting in the $\hat{S}_0\hat{T}_0$ sub-space, $\hat{H}^{T_{+1}T_{-1}}$ acts in the $\hat{T}_{+1}\hat{T}_{-1}$ sub-space, and $\hat{H}_{\mu w}$ is the microwave operator (see Supporting Information 2 for further definitions). With this Hamiltonian, Liouville-space time domain calculations were performed based on the equation of motion:

$$\frac{\partial}{\partial t}\hat{\rho}(t) = -i\hat{\hat{L}}\hat{\rho}(t), \qquad \hat{\hat{L}} = \hat{\hat{H}}_{rot} + i\hat{\hat{R}}, \qquad O(t) = \left\langle \hat{O} \middle| \hat{\rho}(t) \right\rangle \qquad (3)$$

where $\hat{\rho}(t)$ is the state vector of the system, and $\hat{\hat{L}}$ is the Liouvillian containing the rotating frame Hamiltonian superoperator plus the relaxation superoperator $\hat{\hat{R}}$ accounting for the stochastic rotation of all anisotropies. The latter was computed according to the Bloch-Redfield-Wangsness relaxation theory[22,36] both by analytical[24] and numerical[25,26] means, including all possible longitudinal, transverse and cross-correlated pathways. Since the spin system was considered at room temperature, Di-Bari - Levitt thermalization was used.[37] Considering that the exchange coupling has the same order of magnitude as the electron Larmor frequency, these calculations incorporated into the relaxation superoperator a scalar relaxation of the first kind.[38] This did not have an effect on the DNP enhancement, since in the $\omega_{\Delta e} \to 0$ case in question, the exchange coupling, $\vec{\hat{E}}_1 \cdot \vec{\hat{E}}_2$, commutes with the Zeeman interaction, $\hat{E}_{1Z} + \hat{E}_{2Z}$. Time evolution $\hat{O}(t)$ of multiple spin state populations was calculated by taking their scalar products with the state vector $\hat{\rho}(t)$ at each time.

## 3. Features of J-DNP

Figure 2 shows how the isotropic exchange coupling modulates the maximum nuclear enhancement achievable by J-DNP upon on-resonance irradiation at the electron Larmor frequency of the biradicals, as a function of magnetic field $B_0$. These numerically simulated plots[39] predict that, for every field strength, there are two $J_{ex}= \pm(\omega_E + \omega_N)$ values for which microwave irradiation leads to a nuclear enhancement close to the maximum $\gamma_E/2\gamma_N$ achievable value.

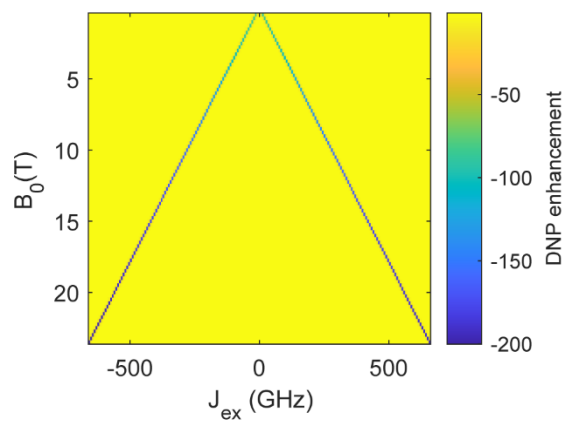

**Fig. 2:** Simulated DNP enhancement achieved within 20 ms of microwave irradiation as a function of $J_{ex}$ and $B_0$. The plots arise from time domain simulations using the parameters in Table 1, a biradical/proton dipolar-coupled triad $\tau_c$ = 500 ps, and an on-resonance irradiation at the electron Larmor frequency of the biradicals. In this and other graphs shown below, DNP enhancements denote the achieved nuclear polarization, normalized by its Boltzmann counterpart at the same temperature and field.

Unlike Overhauser DNP and CCDNP enhancements,[18] where nuclear polarization enhancements are observed at the steady state, the nuclear polarization enhancements shown in Figure 2 and arising at the $J_{ex}= \pm(\omega_E+\omega_N)$ conditions, are transient phenomena. This is illustrated in Figure 3, which shows the enhancement's time-dependence for different rotational correlation times and magnetic fields, for spin systems with an optimally chosen $J_{ex}=+(\omega_E+\omega_N)$ coupling. These graphs show a nuclear polarization that rapidly builds up and then decays into steady states which, at the usual NMR fields used in analytical/biophysical studies, are essentially the Boltzmann equilibrium nuclear magnetization. However, in all cases, substantial transient enhancements are observed after 10-50 ms of continuous microwave irradiation, with precise timings and maximal values that depend on $B_0$ and on the correlation time $\tau_c$ of the biradical/proton triad. In fact, notice that these transient enhancements improve slightly with both higher $B_0$s and slower $\tau_c$s of the biradical/proton triad. Additional differences between the behaviours of Overhauser and J-DNP are discussed in Fig. S3 of Supporting Information 1, which compares expectations from a two-spin proton/electron system (Overhauser DNP), and the changes arising when a second electron is added to form a three-spin proton/biradical system, where the two electrons interact with $J_{ex}=+(\omega_E+\omega_N)$. Upon introducing such second electron, weak steady-state nuclear polarization values similar to those arising in Overhauser DNP are reached – but on the way to those steady states, there are strong transient enhancements.

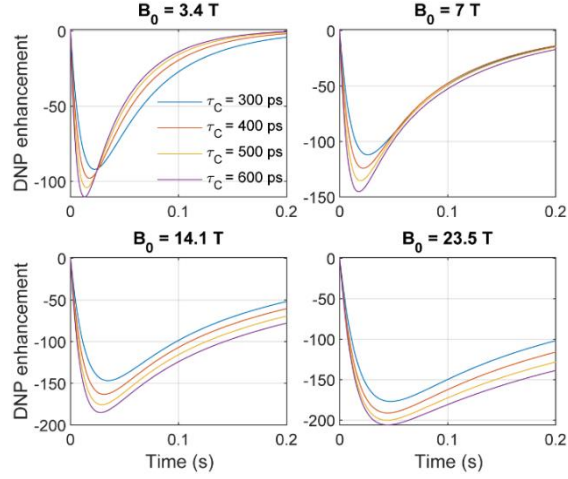

**Fig. 3:** Time domain simulations showing the evolution of the DNP enhancement observed under continuous microwave irradiation of the electrons, for an array of $B_0$ and of $\tau_c$ values of the biradical/proton dipolar-coupled triad. For all fields $J_{ex}$ was set to $+(\omega_E + \omega_N)$; other simulation parameters are as given in Table 1.

### 4. The physics of J-DNP

The physics that drives the J-DNP effects summarized in Figures 2-5, is reminiscent of the cross-correlation CIDNP mechanism. According to the radical pair theory,[28,29,40] nuclear magnetization enhancement in CIDNP proceeds from three processes: (i) singlet/triplet interconversion within a spin-correlated biradical; (ii) a modulation of the rate of this interconversion by hyperfine couplings (i.e., different fates of the electron states depending on whether the nuclear spin state is $\alpha$ or $\beta$); and (iii) a rapid nuclear spin relaxation of the unreacted triplet biradical, acting as a nuclear spin state filter. In the J-DNP case, the biradical is not (photo)chemically produced and does not recombine after a transient action; still, the singlet-triplet behaviour vis-à-vis the nuclear spin once again becomes relevant. In the J-DNP case, it is microwaves (rather than a laser) that drive the system away from the thermal equilibrium. As in CIDNP, it is essential that the nucleus is differentially hyperfine-coupled to the two electrons for the J-DNP enhancement to occur (the effect disappears otherwise, see Supporting information 6). It is then the different relaxation characteristics of the two-electron singlet/triplet states when facing the nuclear $\alpha$ or $\beta$ states, that build up the nuclear polarization. Figures 4a-4d further clarify this, by showing the time dependencies of the three-spin population operators $\hat{O}_{\alpha/\beta}$,[41,42] computed from direct-product of two-electron triplet/singlet states with a nuclear spin state in Zeeman basis, that can be in either the $\alpha$ or $\beta$ state. These states are represented as $\hat{T}_{\pm 1,\alpha/\beta}$, $\hat{T}_{0,\alpha/\beta}$ and $\hat{S}_{0,\alpha/\beta}$ (see Supporting Information 3 for the expressions of these $\hat{O}_{\alpha/\beta}$ as Cartesian operators). Figure 4e shows the states arising from the difference between $\hat{O}_\alpha$ and $\hat{O}_\beta$, defined as:

$$\hat{O}^{(e1e2)}\hat{N}_Z = \left|\hat{O}^{(e1e2)}\right\rangle\left\langle\hat{O}^{(e1e2)}\right|\hat{N}_Z = \frac{1}{2}\left(\hat{O}_\alpha - \hat{O}_\beta\right) \qquad (4)$$

where $\hat{O}^{(e1e2)}$ denotes the two-electron singlet and triplet, and $\hat{N}_Z$ is the longitudinal nuclear magnetization. Figure 4f shows the overall $\hat{N}_Z$ amplitude calculated upon summing the amplitudes of $\hat{S}_0\hat{N}_Z$, $\hat{T}_0\hat{N}_Z$ and $\hat{T}_{\pm 1}\hat{N}_Z$, normalized to the equilibrium nuclear magnetization.

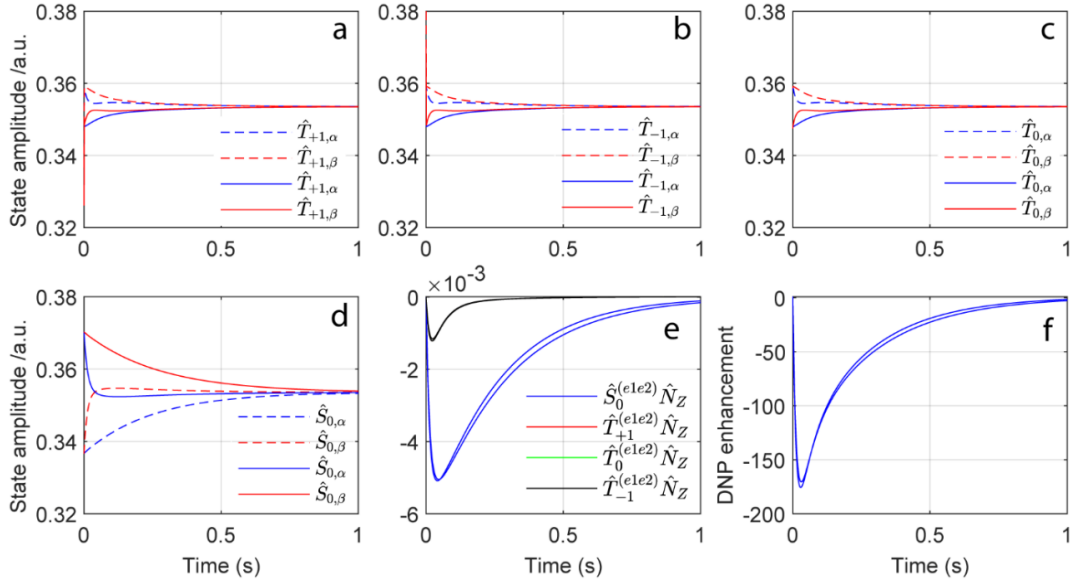

**Fig. 4:** Time evolution of the three-spin population operators: $\hat{T}_{\pm 1,\alpha/\beta}$, $\hat{T}_{0,\alpha/\beta}$ and $\hat{S}_{0,\alpha/\beta}$ (a-d), and of the $\hat{N}_Z\hat{S}_0$, $\hat{N}_Z T_{\pm 1}$ and $\hat{N}_Z\hat{T}_0$ states (e) arising from the difference between the various $\hat{O}_\alpha$ and $\hat{O}_\beta$. Time evolution of the overall longitudinal nuclear magnetization $\hat{N}_Z$ (f), showing the sum of all $\hat{N}_Z$-related contributions, normalized by the thermal nuclear magnetization under the same conditions. Plots (a-d) are normalized to the total electron spin population of one, as defined in Supporting Information 3. $J_{ex}= +(\omega_E+\omega_N)$ was used for the dashed lines, and $J_{ex} = -(\omega_E+\omega_N)$ was used for the continuous lines. In (e), the red and green traces are barely seen as they fall underneath an overlapping black trace. The slightly different curves in (f) reflect $J_{ex}= \pm(\omega_E+\omega_N)$ conditions, respectively. For all these calculations $B_0$ = 14.08 T, $\tau_c$ = 500 ps for the biradical/proton dipolar-coupled triad; all other parameters as given in Table 1.

At time zero, the system is at thermal equilibrium, and negligible differences arise between the populations of the α and β nuclear states – even if singlet and triplet electron state populations differ. When the microwave irradiation is turned on, the electron spin is taken out of the thermal equilibrium and electron saturation sets in, the triplet states are mixed, and their populations start to converge to similar amplitudes (Figures 4a-4c). However, at the $J_{ex}= \pm(\omega_E+\omega_N)$ condition, the rates at which electron states settle into this new equilibrium are different for α and β nuclear components: when $J_{ex}= +(\omega_E+\omega_N)$ the α component of the triplets (Fig. 4, blue dashed line) reaches the steady state faster than the β component (red dashed line), while the β component of the singlet (red dashed line) reaches steady state faster than the α component of the singlet (blue dashed line).[43] The opposite α/β behaviour arises when $J_{ex}= -(\omega_E+\omega_N)$. Still, because different initial conditions exist when $J_{ex}= \pm(\omega_E+\omega_N)$, both situations lead to similar transient nuclear polarization build-ups, as shown in Figure 4e. Consequently, due to the different self-relaxation rates of the α and β nuclear components of the triplet and singlet electron states, a sizable nuclear polarization $\hat{N}_Z$ builds up. As can be appreciated in Figure 4f, this polarization is only weakly dependent on the sign of $J_{ex}$. Also note that, as eventually all electron/nuclear states reach the same populations, this J-DNP build up is transient: no nuclear polarization gains are predicted when the steady-state conditions examined in conventional Overhauser DNP analyses, are considered.

The plots shown in Figure 4 were computed assuming continuous microwave irradiation. To avoid the decay of the enhancement – dominated by the nuclear $T_1$ decay draining the $\hat{N}_Z\hat{S}_0$ state – microwaves could be turned off when nuclear magnetization has reached a maximum; the electron population operators would then relax back to thermal equilibrium, and the build-up could be "pumped" again by repeated irradiation. However, in an actual biradical-driven enhancement experiment, the polarizing proton would not be covalently bound; rather, it would be randomly diffusing at rates of ca. 1 μm/ms. The interaction with the polarizing biradical would thus be short-lived, and the proton would get repeatedly polarized by different transient encounters with different biradicals; a certain "pulsing" of the effect will therefore occur spontaneously.[13,44,45]

## 5. Analysis of the self-relaxation rates driving J-DNP

Figure 5 examines another facet of J-DNP, by showing how the self-relaxation rates of the $\hat{T}_{\pm 1,\alpha/\beta}$, $\hat{T}_{0,\alpha/\beta}$ and $\hat{S}_{0,\alpha/\beta}$ population operators driving the population decays of the various nuclear spin states, change over a range of magnetic fields $B_o$ and exchange couplings $J_{ex}$. Notice the marked differentials arising whenever $J_{ex}$ matches $\pm(\omega_E + \omega_N)$, between the self-relaxation rates of various $\hat{O}_\alpha$ and $\hat{O}_\beta$ operators. It is these differential decays that drive the enhancements shown in Figs. 1 – 4.

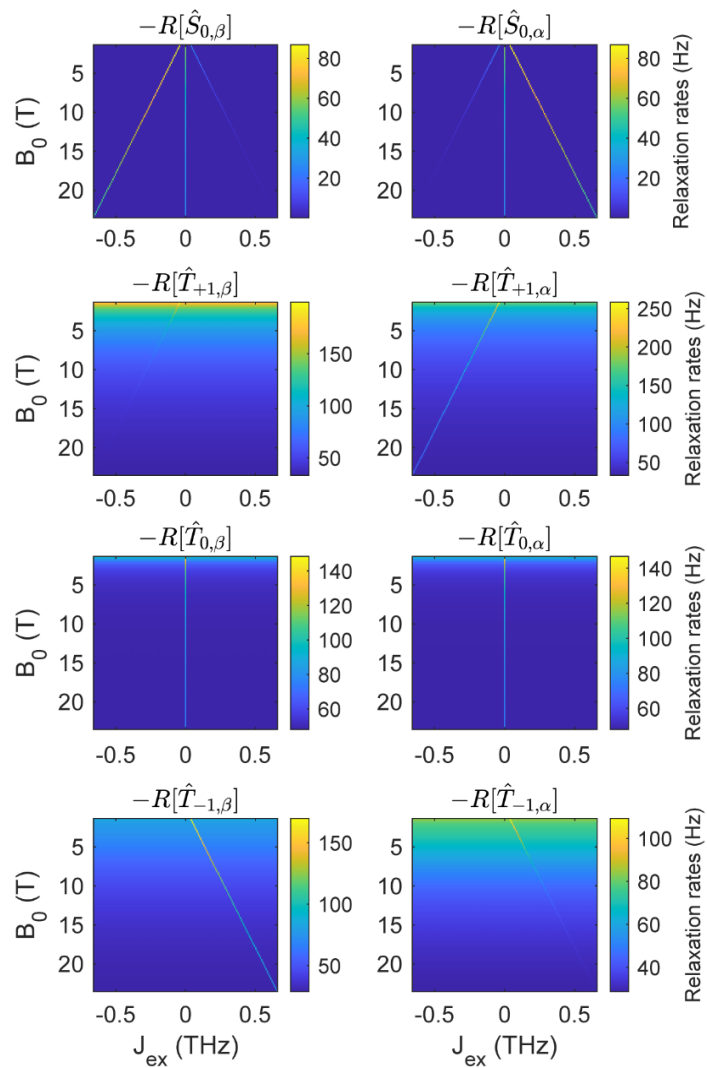

**Fig. 5:** Numerically computed (Redfield theory) self-relaxation rates of the three-spin population operators as a function of $J_{ex}$ and $B_0$. Rates of $\hat{S}_{0,\beta}$, $\hat{T}_{0,\beta}$ and $\hat{T}_{\pm 1,\beta}$ are on the left, and rates of $\hat{S}_{0,\alpha}$, $\hat{T}_{0,\alpha}$ and $\hat{T}_{\pm 1,\alpha}$ are on the right. Calculations relied on the parameters of Table 1 and on $\tau_c$ = 500 ps for the biradical/proton dipolar-coupled triad.

The numerical trends in Figure 5 can be explained by considering the analytical expressions for the relaxation rates of singlet and the triplet states, as derived from Redfield's relaxation theory.[24,36] As long as this theoretical model is valid ($\tau_c \leq 2$ ns), these rates will be given by sums of terms involving products of second rank norms squared and/or other quadratic products,[46] times spectral density functions. For the $\hat{S}_{0,\alpha/\beta}$ operators these rates can be summarized as:

$$-R\left[\hat{S}_{0,\beta}\right] = \frac{\Delta_{\mathrm{AHF}}^2}{180} J\left(J_{\mathrm{ex}} - \omega_{\mathrm{E}} + \omega_{\mathrm{N}}\right) + \frac{\Delta_{\mathrm{AHF}}^2}{30} J\left(J_{\mathrm{ex}} + \omega_{\mathrm{E}} + \omega_{\mathrm{N}}\right) +$$
$$+ \frac{\left[\Delta_{\mathrm{AHF}}^2 + 4\aleph_{\Delta G,\Delta G\text{-}\Delta HF}\right]}{120} J\left(J_{\mathrm{ex}} - \omega_{\mathrm{E}}\right) + \frac{\left[\Delta_{\mathrm{AHF}}^2 + 4\aleph_{\Delta G,\Delta G\text{-}\Delta HF}\right]}{120} J\left(J_{\mathrm{ex}} + \omega_{\mathrm{E}}\right) + \ldots \quad (5)$$

and

$$-R\left[\hat{S}_{0,\alpha}\right] = \frac{\Delta_{\mathrm{AHF}}^2}{180} J\left(J_{\mathrm{ex}} + \omega_{\mathrm{E}} - \omega_{\mathrm{N}}\right) + \frac{\Delta_{\mathrm{AHF}}^2}{30} J\left(J_{\mathrm{ex}} - \omega_{\mathrm{E}} - \omega_{\mathrm{N}}\right) +$$
$$+ \frac{\left[\Delta_{\mathrm{AHF}}^2 + 4\aleph_{\Delta G,\Delta G+\Delta HF}\right]}{120} J\left(J_{\mathrm{ex}} - \omega_{\mathrm{E}}\right) + \frac{\left[\Delta_{\mathrm{AHF}}^2 + 4\aleph_{\Delta G,\Delta G+\Delta HF}\right]}{120} J\left(J_{\mathrm{ex}} + \omega_{\mathrm{E}}\right) + \ldots \quad (6)$$

For the $\hat{T}_{+1,\alpha/\beta}$ operators, the self-relaxation rates are:

$$-R\left[\hat{T}_{+1,\beta}\right] = \frac{\Delta_{\mathrm{AHF}}^2}{180} J\left(J_{\mathrm{ex}} + \omega_{\mathrm{E}} - \omega_{\mathrm{N}}\right) + \frac{\left[\Delta_{\mathrm{AHF}}^2 + 4\aleph_{\Delta G,\Delta G\text{-}\Delta HF}\right]}{120} J\left(J_{\mathrm{ex}} + \omega_{\mathrm{E}}\right) + \ldots \quad (7)$$

and

$$-R\left[\hat{T}_{+1,\alpha}\right] = \frac{\Delta_{\mathrm{AHF}}^2}{30} J\left(J_{\mathrm{ex}} + \omega_{\mathrm{E}} + \omega_{\mathrm{N}}\right) + \frac{\left[\Delta_{\mathrm{AHF}}^2 + 4\aleph_{\Delta G,\Delta G+\Delta HF}\right]}{120} J\left(J_{\mathrm{ex}} + \omega_{\mathrm{E}}\right) + \ldots \quad (8)$$

For the $\hat{T}_{0,\alpha/\beta}$ operators, the self-relaxation rates are:

$$-R\left[\hat{T}_{0,\beta}\right] = \frac{\Delta_{\mathrm{CSA}}^2}{15} J\left(\omega_{\mathrm{N}}\right) + \ldots \quad (9)$$

and

$$-R\left[\hat{T}_{0,\alpha}\right] = \frac{\Delta_{\mathrm{CSA}}^2}{15} J\left(\omega_{\mathrm{N}}\right) + \ldots \quad (10)$$

and for $\hat{T}_{-1,\alpha/\beta}$ operators, the self-relaxation rates:

$$-R\left[\hat{T}_{-1,\beta}\right] = \frac{\Delta_{\mathrm{AHF}}^2}{30} J\left(J_{\mathrm{ex}} - \omega_{\mathrm{E}} - \omega_{\mathrm{N}}\right) + \frac{\left[\Delta_{\mathrm{AHF}}^2 + 4\aleph_{\Delta G,\Delta G\text{-}\Delta HF}\right]}{120} J\left(J_{\mathrm{ex}} - \omega_{\mathrm{E}}\right) + \ldots \quad (11)$$

and

$$-R\left[\hat{T}_{-1,\alpha}\right] = \frac{\Delta_{\mathrm{AHF}}^2}{180} J\left(J_{\mathrm{ex}} - \omega_{\mathrm{E}} + \omega_{\mathrm{N}}\right) + \frac{\left[\Delta_{\mathrm{AHF}}^2 + 4\aleph_{\Delta G,\Delta G+\Delta HF}\right]}{120} J\left(J_{\mathrm{ex}} - \omega_{\mathrm{E}}\right) + \ldots \quad (12)$$

where **ΔG** = **G₁**-**G₂** and **ΔHF** = **HFC₁**-**HFC₂** are anisotropies associated to the differences between the two *g*- and electron/nuclear hyperfine coupling tensors, respectively; **CSA** is the chemical shift anisotropy tensor; and, as is usual in spin relaxation theory,[22,23] all rates contain combinations of second-rank norms squared $\Delta_A^2$ for all the aforementioned tensors **A**,[46] and second-rank scalar products $\aleph_{A,B}$ of tensors **A** and **B**. The "…" in Eqs. (5)-(12) denote terms that contribute equally to $\hat{O}_\alpha$ and $\hat{O}_\beta$ and thus are not relevant for the J-DNP effect; full analytical expressions for all the self-relaxation rates are given in Supporting Information 4, and the definition of second rank norm squared and scalar product is given in Supporting Information 5.

As mentioned for the Overhauser DNP, at high B₀ fields, the contribution of terms $J(\omega_E) \approx 0$, to the rates in Eqs. (5) – (12), can be disregarded. Exceptions, however, will arise when a spectral density's ω-argument is $\omega_N$, or when it can be made zero by $J_{ex}$ - such spectral density functions will no longer be negligible and the terms cannot be disregarded. The inspection of Eqs. (5) – (12) reveal many such potential terms, that as a result of them, large differences will arise between the self-relaxation rates of the $\hat{O}_\alpha$ and $\hat{O}_\beta$ population operators under $J_{ex} \approx \pm\omega_E$ conditions, leading to the transient generation of non-zero $\hat{N}_Z\hat{S}_0$, $\hat{N}_Z\hat{T}_{\pm 1}$ states shown in Fig. 4. This can be most easily appreciated for $\hat{S}_{0,\alpha}$ and $\hat{S}_{0,\beta}$ relaxation rates: the latter will be dominated by the $J(J_{ex}+\omega_E+\omega_N)$ term if $J_{ex}= -(\omega_E +\omega_N)$, and the former by the $J(J_{ex} -\omega_E -\omega_N)$ if $J_{ex}= +(\omega_E+\omega_N)$. In either case, such $J_{ex}$ condition will result in large differences between $R[\hat{O}_\alpha]$ and $R[\hat{O}_\beta]$ due to the cancellation of $\omega_E$ by $J_{ex}$. Similar differences can be observed in Figure 5 also for $\hat{T}_{\pm 1,\alpha}$ and $\hat{T}_{\pm 1,\beta}$. Supporting Information 4 provides additional information about these self-relaxation rates. Note as well that, even though the absolute values of these rates decrease with $B_0$ and $\tau_c$, their differences actually remain present (Figures S4 and S5, Supporting information), and still lead to sizable longitudinal nuclear magnetizations; this helps to understand the increased efficiencies with $B_0$ and $\tau_c$ shown for J-DNP enhancements in Figures 1 - 4.

It is also instructive to consider the maximum enhancement expected from the differences between the $R[\hat{O}_\alpha]$ and the $R[\hat{O}_\beta]$ relaxation rates. Assuming for simplicity that one of the states relaxes infinitely fast while the other has no relaxation, an assumption of complete microwave-driven electron saturation will lead to a nuclear polarization reaching a 0.5 value – i.e., the NMR signal would be enhanced by $|\gamma_E/2\gamma_N|$. The data in Figures 1-4 show $\hat{N}_Z$ enhanced to a significant fraction of this upper bound.

## 6. Conclusions and outlook

The present study introduced a new proposal for enhancing signals in solution state NMR, that could potentially provide substantial sensitivity gains for a wide range of magnetic fields and rotational correlation times. This J-DNP mechanism uses biradicals, it is transient, and it emerges in the hitherto unexplored $J_{ex}= \pm(\omega_E+\omega_N)$ regimes. This requires exchange couplings in the order of the electron Larmor frequencies; i.e., ranging between ≈200-700 GHz for NMR measurements in 7-23.5 T fields. Such $J_{ex}$ values are not out of the ordinary: radical monomers connected by conjugated linkers having inter-electron exchange coupling in this order, have been reported in the literature.[19-21] Under such conditions, both numerical and analytical simulations provide coinciding predictions of significant NMR polarization build-ups. These enhancements require irradiation times lasting a fraction of a second and, if implemented at the right electron Larmor frequency, they can proceed efficiently with microwave nutation fields ≤ 1 MHz (see Supporting information 1 for additional details). Supporting Information 6 demonstrates that similar enhancements are reached with zero *g*- and shielding tensor anisotropies; the enhancements are also preserved if the two electrons have different but isotropic *g*-tensors. Much like the radical pair mechanism of CIDNP, the mechanism of J-DNP requires different hyperfine coupling to the two electrons; this drives nuclear state dependent electron relaxation which disappears if the nucleus is symmetrically placed between the two electrons. The enhancement is also quenched if the two electrons have different anisotropic *g*-tensors – the resulting ΔG-driven electron relaxation would then, at high magnetic fields, overtake the weaker differential relaxation mechanism arising different hyperfine coupling to the two electrons. Likewise, the enhancement goes to zero if the nucleus remains too distant from the biradical: for instance, a proton placed 20 Å away from the biradical may require over 1 s to achieve significant polarization gains – a time by which the DNP effect will lose against competing relaxation pathways.

Despite its encouraging conclusions, the present study also made a number of strong assumptions. It assumes that the non-Redfield relaxation terms in monoradicals resembled those acting in biradicals (Supporting Information 1); it remains to be seen how realistic this assumption is. Another major approximation was assuming a fixed electron/electron/nuclear geometry over the course of the DNP process: as the latter requires 10-100 ms and molecules in regular liquids diffuse tens of microns over such times, this fixed molecular geometry assumption is clearly unrealistic. On the other hand, a similar fixed-geometry assumption is successfully used and leads to realistic predictions in Overhauser DNP, where it works by virtue of DNP's independence on maintaining specific cross-correlations (a requirement present in our previous CCDNP proposal). Likewise, we hypothesize that the different $R[\hat{O}_{\alpha/\beta}]$ rates required for J-DNP will be preserved regardless of the transient contact that a specific nucleus makes with an ensemble of biradicals over the course of its spin polarization process. Experiments are in progress to evaluate the correctness of these assumptions.

## 7. Acknowledgments

This project was funded by the Weizmann-UK Joint Research Programme, the Israel Science Foundation (ISF 965/18), the Perlman Family Foundation, and the US National Science Foundation (grants numbers CHE-1808660, DMR-1644779). MGC acknowledges Weizmann's Faculty of Chemistry for a Dean Fellowship. LF holds the Bertha and Isadore Gudelsky Professorial Chair and Heads the Clore Institute for High-Field Magnetic Resonance Imaging and Spectroscopy, whose support is acknowledged.

## 8. Author contribution

LF and MGC discovered the J-DNP effect and, jointly with IK, provided a theoretical description of its mechanism. IK has written Spinach kernel code and the symbolic processing engine, MGC wrote the simulation scripts to perform analytical and numerical simulations, and derived the analytical expressions of the relaxation rates. All authors discussed and wrote the paper.

## 9. Competing interests

The authors declare no competing interests.

## 10. References

1  Overhauser, A. W. Polarization of nuclei in metals. *Phys. Rev.* **92**, 411 (1953).
2  Carver, T. R. & Slichter, C. P. Experimental verification of the Overhauser nuclear polarization effect. *Phys. Rev.* **102**, 975 (1956).
3  Hu, K.-N., Yu, H.-h., Swager, T. M. & Griffin, R. G. *J. Am. Chem. Soc.* **126**, 10844 (2004).
4  Bajaj, V. S. *et al.* 250 GHz CW gyrotron oscillator for dynamic nuclear polarization in biological solid state NMR. *J. Magn. Reson.* **189**, 251 (2007).
5  Bajaj, V. *et al.* Dynamic nuclear polarization at 9 T using a novel 250 GHz gyrotron microwave source. *J. Magn. Reson.* **213**, 404 (2011).
6  Hu, K. N., Debelouchina, G. T., Smith, A. A. & Griffin, R. G. Quantum mechanical theory of dynamic nuclear polarization in solid dielectrics. *J. Chem. Phys.* **134**, 125105 (2011).
7  Ardenkjaer-Larsen, J. H. *et al.* Increase in signal-to-noise ratio of > 10,000 times in liquid-state NMR. *Proc. Natl. Acad. Sci. U.S.A.* **100**, 10158 (2003).
8  Golman, K., Lerche, M., Pehrson, R. & Ardenkjaer-Larsen, J. H. Metabolic imaging by hyperpolarized 13C magnetic resonance imaging for in vivo tumor diagnosis. *Cancer Res.* **66**, 10855 (2006).

9   Loening, N. M., Rosay, M., Weis, V. & Griffin, R. G. Solution-state dynamic nuclear polarization at high magnetic field. *J. Am. Chem. Soc.* **124**, 8808 (2002).

10  Liu, G. Q. *et al.* One-thousand-fold enhancement of high field liquid nuclear magnetic resonance signals at room temperature. *Nature Chemistry* **9**, 676 (2017).

11  Dubroca, T., Wi, S., van Tol, J., Frydman, L. & Hill, S. Large volume liquid state scalar Overhauser dynamic nuclear polarization at high magnetic field. *PCCP* **21**, 21200 (2019).

12  Parigi, G., Ravera, E., Bennati, M. & Luchinat, C. Understanding Overhauser Dynamic Nuclear Polarisation through NMR relaxometry. *Mol. Phys.* **117**, 888 (2019).

13  Levien, M., Hiller, M., Tkach, I., Bennati, M. & Orlando, T. Nitroxide Derivatives for Dynamic Nuclear Polarization in Liquids: The Role of Rotational Diffusion. *J. Phys. Chem. Lett.* **11**, 1629 (2020).

14  Hausser, K. H. & Stehlik, D. *Dynamic nuclear polarization in liquids, in Advances in Magnetic and Optical Resonance*. Vol. 3, p.79-139 (Elsevier, 1968).

15  Griesinger, C. *et al.* Dynamic nuclear polarization at high magnetic fields in liquids. *Prog. Nucl. Magn. Reson. Spectrosc.* **64**, 4 (2012).

16  Prisner, T., Denysenkov, V. & Sezer, D. Liquid state DNP at high magnetic fields: Instrumentation, experimental results and atomistic modelling by molecular dynamics simulations. *J. Magn. Reson.* **264**, 68 (2016).

17  Ravera, E., Luchinat, C. & Parigi, G. Basic facts and perspectives of Overhauser DNP NMR. *J. Magn. Reson.* **264**, 78 (2016).

18  Concilio, M. G., Soundararajan, M., Frydman, L. & Kuprov, I. High-field solution state DNP using cross-correlations. *J. Magn. Reson.* **326**, 106940 (2021).

19  Reginsson, G. W., Kunjir, N. C., Sigurdsson, S. T. & Schiemann, O. Trityl radicals: spin labels for nanometer-distance measurements. *Chem. Eur. J.* **18**, 13580 (2012).

20  Jassoy, J. J., Meyer, A., Spicher, S., Wuebben, C. & Schiemann, O. Synthesis of Nanometer Sized Bis- and Tris-trityl Model Compounds with Different Extent of Spin-Spin Coupling. *Molecules* **23**, 682 (2018).

21  Fleck, N. *et al.* C-C Cross-Coupling Reactions of Trityl Radicals: Spin Density Delocalization, Exchange Coupling, and a Spin Label. *J. Org. Chem.* **84**, 3293 (2019).

22  Redfield, A. *The theory of relaxation processes, in Advances in Magnetic and Optical Resonance*. Vol. 1, p.1-32 (Elsevier, 1965).

23  Goldman, M. Formal theory of spin--lattice relaxation. *J. Magn. Reson.* **149**, 160 (2001).

24  Kuprov, I., Wagner-Rundell, N. & Hore, P. Bloch-Redfield-Wangsness theory engine implementation using symbolic processing software. *J. Magn. Reson.* **184**, 196 (2007).

25  Kuprov, I. Diagonalization-free implementation of spin relaxation theory for large spin systems. *J. Magn. Reson.* **209**, 31 (2011).

26  Goodwin, D. & Kuprov, I. Auxiliary matrix formalism for interaction representation transformations, optimal control, and spin relaxation theories. *J. Chem. Phys.* **143**, 084113 (2015).

27  Bargon, J., Fischer, H. & Johnsen, U. Kernresonanz-Emissionslinien während rascher Radikalreaktionen. *Zeitschrift für Naturforschung A* **22**, 1551 (1967).

28  Kaptein, R. & Oosterhoff, L. J. Chemically induced dynamic nuclear polarization II: (Relation with anomalous ESR spectra). *Chem. Phys. Lett.* **4**, 195 (1969).

29  Closs, G. L. Mechanism explaining nuclear spin polarizations in radical combination reactions. *J. Am. Chem. Soc.* **93**, 1546 (1971).

30  Santabarbara, S. *et al.* Bidirectional electron transfer in photosystem I: determination of two distances between P700+ and A1- in spin-correlated radical pairs. *Biochemistry* **44**, 2119 (2005).

31  Ardenkjaer-Larsen, J. H. *et al.* EPR and DNP properties of certain novel single electron contrast agents intended for oximetric imaging. *J. Magn. Reson.* **133**, 1 (1998).

32  Wind, R. A. & Ardenkjaer-Larsen, J. H. 1H DNP at 1.4 T of water doped with a triarylmethyl-based radical. *J. Magn. Reson.* **141**, 347 (1999).

33  Hofer, P. *et al.* Field dependent dynamic nuclear polarization with radicals in aqueous solution. *J. Am. Chem. Soc.* **130**, 3254 (2008).

34  Yong, L. *et al.* Electron spin relaxation of triarylmethyl radicals in fluid solution. *J. Magn. Reson.* **152**, 156 (2001).

35  Owenius, R., Eaton, G. R. & Eaton, S. S. Frequency (250 MHz to 9.2 GHz) and viscosity dependence of electron spin relaxation of triarylmethyl radicals at room temperature. *J. Magn. Reson.* **172**, 168 (2005).

36  Wangsness, R. K. & Bloch, F. The dynamical theory of nuclear induction. *Phys. Rev.* **89**, 728 (1953).

37  Levitt, M. H. & Di Bari, L. Steady state in magnetic resonance pulse experiments. *Phys. Rev. Lett.* **69**, 3124 (1992).

38  Kuprov, I. *et al.* Anomalous Nuclear Overhauser Effects in Carbon-Substituted Aziridines: Scalar Cross-Relaxation of the First Kind. *Angew. Chem. Int. Ed.* **54**, 3697 (2015).

39  Hogben, H., Krzystyniak, M., Charnock, G., Hore, P. & Kuprov, I. Spinach - a software library for simulation of spin dynamics in large spin systems. *J. Magn. Reson.* **208**, 179 (2011).

40  Hore, P. J. & Broadhurst, R. W. Photo-CIDNP of biopolymers. *Prog. NMR Spectrosc.* **25**, 345 (1993).

41  Vega, S. Fictitious spin 1/2 operator formalism for multiple quantum NMR. *J. Chem. Phys.* **68**, 5518 (1978).

42  Ernst, R. R., Bodenhausen, G. & Wokaun, A. *Principles of Nuclear Magnetic Resonance in One and Two Dimensions*. p.610 (Oxford: Clarendon Press, 1987).

43  Kuprov, I., Craggs, T. D., Jackson, S. E. & Hore, P. J. Spin relaxation effects in photochemically induced dynamic nuclear polarization spectroscopy of nuclei with strongly anisotropic hyperfine couplings. *J. Am. Chem. Soc.* **129**, 9004 (2007).

44  Müller-Warmuth, W., Vilhjalmsson, R., Gerlof, P., J., S. & Trommel, J. Intermolecular interactions of benzene and carbon tetrachloride with selected free radicals in solution as studied by 13C and 1H dynamic nuclear polarization. *Mol. Phys.* **31**, 1055 (1976).

45  Levien, M. *et al.* Spin density localization and accessibility of organic radicals affect liquid-state DNP efficiency. *PCCP* **23**, 4480 (2021).

46  Blicharski, J. Nuclear magnetic relaxation by anisotropy of the chemical shift. *Z Naturforsch Pt A* **27**, 1456 (1972).

Supplementary Information for

# *J*-driven Dynamic Nuclear Polarization for sensitizing high field solution state NMR

*Maria Grazia Concilio[1*], Ilya Kuprov[2], Lucio Frydman,[1,3*]*

[1]Department of Chemical and Biological Physics, Weizmann Institute of Science, Rehovot, Israel.
[2]School of Chemistry, University of Southampton, Southampton, UK.
[3]National High Magnetic Field Laboratory, Tallahassee, Florida, USA.

**Supporting Information 1: $R_1$/$R_2$ relaxation rate analysis for mono- and bis-trityl radicals**

As this is a purely theoretical study, we consider it necessary to test our numerical simulations against experimental measurements. Given the paucity of solution-state NMR data about relaxation rates in biradicals, theoretical models were used to estimate electron longitudinal and transverse relaxation rates for trityl monoradicals – for which solution-state measurements have been reported in [34,35]. Following the treatment in [35], electron $R_1$ rate for this radical will be assumed to have three contributions:

$$R_{1,\text{total}} = R_{1,\text{local}} + R_{1,\text{BRW}} + R_{1,\text{solvent}}$$

$$\approx R_{1,\text{local}} + R_{1,\text{BRW}} \qquad \text{(S1a)}$$

Here $R_{1,\text{local}}$ is the self-relaxation rate arising from the local vibrational modes related to the stretching of bonds in the radical, and is a magnetic field independent contribution estimated by Eaton et al [35] as 5.9 x 10$^4$ Hz for per-deutero and 6.1 x 10$^4$ Hz for per-protio trityls. The Redfield component:

$$R_{1,\text{BRW}} = n\left[\left(\frac{\Delta_{\text{HF,intra}}^2}{9}\right)\left(\frac{\tau_C}{1+\tau_C^2\omega_E^2}\right)\right]_{\text{hyperfine relaxation}} + \underbrace{\left(\frac{2\Delta_G^2}{15}\right)\left(\frac{\tau_C}{1+\tau_C^2\omega_E^2}\right)}_{\text{g-anisotropy relaxation}} \qquad \text{(S1b)}$$

is the relaxation rate which arises from the stochastic modulation of the radical's *g*-tensor anisotropy and of the hyperfine interactions between the electron and the n protons interacting with it within the radical (that in the case of a trityl monomer can be considered approximately the same); this is the kind of rate computed in the main text based on Bloch-Redfield-Wangsness theory in terms of $\Delta_G^2$ and $\Delta_{\text{HF,intra}}^2$ coupling strengths; i.e., of the second-rank norm Blicharsky squared,[46] already mentioned in the main text and defined in Eq. S58 in Supporting Information 5, for the *g*-anisotropy and the hyperfine interaction tensors. Finally,

$$R_{1,\text{solvent}} = n_{\text{solv}}\left[\left(\frac{\Delta_{\text{HF,inter}}^2}{9}\right)\frac{\tau_{\text{solv}}}{\left(1+\tau_{\text{solv}}^2\omega_E^2\right)}\right] \qquad \text{(S1c)}$$

is an *ad hoc* relaxation term, arising from the dipolar interaction between the electron and $n_{\text{solv}}$ solvent molecules; the coupling strength of this relaxation is given by the second-rank norm squared of the dipolar interaction tensor between the electron and the solvent protons $\Delta_{\text{HF,inter}}^2$, and by a correlation time $\tau_{\text{solvent}}$. Eaton et al in [35] have estimated the $R_{1,\text{solvent}}$ in Eq. (S1c) as $\tau_{\text{solv}} = F\,\tau_C$, where F is a ratio between the solvent correlation time and the trityl correlation time, that is different for each radical. As this latter term is much smaller than the remaining relaxation rates it is henceforth ignored in both the mono- and bi-radical estimations. Finally, in a non-viscous solvent, the $R_{2,total}$ corresponds to $R_{1,total} + R_{2,BRW}$ [35]. The BRW theory predicts this second term to be:

$$R_{2,BRW} = n \underbrace{\left[\left(\frac{\Delta_{HF,trityl}^2}{90}\right)\left(2 + \frac{5\tau_C}{(1+\tau_C^2\omega_E^2)} + \frac{3\tau_C}{(1+\tau_C^2\omega_N^2)}\right)\right]}_{\text{hyperfine relaxation}} + \underbrace{\left(\frac{\Delta_G^2}{45}\right)\left(4 + \frac{3\tau_C}{(1+\tau_C^2\omega_E^2)}\right)}_{\text{g-anisotropy relaxation}} \quad (S2)$$

Table S1 compares the $R_{1,total}$ and $R_{2,total}$ values determined from Eqs. (S1a) and (S2), with values reported experimentally in [35] for per-protio and per-deutero trityls at different fields. There is good agreement between both sets. Notice, however, that the low magnetic fields explored in this comparison, it is the local vibrational modes contributing to $R_{1,local}$ that dominate the longitudinal- and transverse relaxation rates. As this parameter was estimated from the experimental data, the good agreement is not surprising.

**Table S1:** Comparison between $T_1$ and $T_2$ calculated using Eq. (S1a) and Eq. (S2) and experimental $T_1$ and $T_2$ in water for trityl-$CD_3$ and trityl-$CH_3$ obtained from [35]. n represents the number of trityl protons dipole-coupled to the radical. For trityl-$CH_3$, the distances between the electron and all 36 protons in the radical were set equal to 5.3 Å.

| Magnetic field / T | $R_1$ from Eq. 1 / Hz | Experimental $R_1$ / Hz | $R_2$ from Eq. 2 / Hz | Experimental $R_2$ / Hz |
|---|---|---|---|---|
| Trityl-$CD_3$ radical, n = 0, $R_{1,local}$= 0.59 x 10$^5$ Hz | | | | |
| 0.33 T (X-band) | 5.9x10$^4$ | 5.9x10$^4$ | 6.7x10$^4$ | 9.1x10$^4$ |
| 0.11 T (S-band) | 5.9x10$^4$ | 6.2x10$^4$ | 6.2x10$^4$ | 8.3x10$^4$ |
| 0.03 T (L-band) | 5.9x10$^4$ | 7.2x10$^4$ | 6.2x10$^4$ | 8.3x10$^4$ |
| 9 x10$^{-3}$ T (250 MHz) | 6.2x10$^4$ | / | 6.2x10$^4$ | 9.1x10$^4$ |
| Trityl-$CH_3$ radical → n = 36, $R_{1,local}$ = 0.61 x 10$^5$ Hz | | | | |
| 0.33 T (X-band) | 6.2x10$^4$ | 6.2x10$^4$ | 2.3x10$^5$ | 1.1x10$^5$ |
| 0.11 T (S-band) | 6.6x10$^4$ | 7.1x10$^4$ | 1.0x10$^5$ | 1.1x10$^5$ |
| 0.03 T (L-band) | 8.3x10$^4$ | 8.3x10$^4$ | 1.1x10$^5$ | 1.1x10$^5$ |
| 9 x10$^{-3}$ T (250 MHz) | 9.1x10$^4$ | / | 1.3x10$^5$ | 1.3x10$^5$ |

It is enlightening to extend this monoradical analysis to the case of trityl bi-radicals. In this case it is reasonable to assume that the contributions coming from solvent-induced and bond-vibration-induced terms to $R_1$ relaxation, will remain similar as those given in Table S1. The electron $R_{1,BRW}$ and $R_{2,BRW}$ longitudinal and transverse rates, however, will be significantly increased by the electron-electron dipolar interaction, and will now be dominated by $J(0)$-containing terms that depend on the rotational correlation time but are independent of the magnetic field. A full analysis of these two terms using symbolic processing software[24] leads to:

$$R_{1,\text{BRW}} = -R\left[\hat{E}_Z\right] = \frac{\Delta_{\text{EE}}^2}{90}J(0) + \frac{\aleph_{(\text{HFC},\Sigma\text{HF}),\text{inter}}}{18}J(\omega_E) +$$

$$+ \frac{\aleph_{(\text{HFC},\Delta\text{HF}),\text{inter}}}{360}\begin{bmatrix}3J(J_{ex}+\omega_E) + 3J(J_{ex}-\omega_E) + 6J(J_{ex}+\omega_E+\omega_N) + \\ +6J(J_{ex}-\omega_E-\omega_N) + J(J_{ex}+\omega_E-\omega_N) + J(J_{ex}-\omega_E+\omega_N)\end{bmatrix} +$$

$$+ n\left(\begin{array}{c}\dfrac{\aleph_{(\text{HFC},\Sigma\text{HF}),\text{intra}}}{18}J(\omega_E) + \\ \dfrac{\aleph_{(\text{HFC},\Delta\text{HF}),\text{intra}}}{360}\begin{bmatrix}3J(J_{ex}+\omega_E) + 3J(J_{ex}-\omega_E) + 6J(J_{ex}+\omega_E+\omega_N) + \\ +6J(J_{ex}-\omega_E-\omega_N) + J(J_{ex}+\omega_E-\omega_N) + J(J_{ex}-\omega_E+\omega_N)\end{bmatrix}\end{array}\right) +$$

$$+ \frac{\left[4\aleph_{G,\Delta G} + 2\aleph_{EE,-\Delta G}\right]}{120}J(J_{ex}-\omega_E) + \frac{\left[4\aleph_{G,\Delta G} + 2\aleph_{EE,\Delta G}\right]}{120}J(J_{ex}+\omega_E) +$$

$$+ \frac{\left[3\Delta_{\text{EE}}^2 + 6\aleph_{G,\Sigma G}\right]}{90}J(\omega_E) + \frac{\Delta_{\text{EE}}^2}{15}J(2\omega_E)$$

(S3)

and

$$R_{2,\text{BRW}} = -R\left[\hat{E}_+\right] = \frac{\Delta_{\text{EE}}^2}{36}J(0) + \frac{4\aleph_{G,\Sigma G}}{90}J(0) +$$

$$+ \frac{\aleph_{(\text{HFC},\Sigma\text{HF}),\text{inter}}}{90}J(0) + \frac{\aleph_{(\text{HFC},\Sigma\text{HF}),\text{inter}}}{120}J(\omega_N) + \frac{\aleph_{(\text{HFC},\Sigma\text{HF}),\text{inter}}}{36}J(\omega_E) +$$

$$+ \frac{\aleph_{(\text{HFC},\Delta\text{HF}),\text{inter}}}{720}\begin{bmatrix}3J(J_{ex}-\omega_E) + 3J(J_{ex}+\omega_E) + 12J(J_{ex}-\omega_E-\omega_N) + 12J(J_{ex}+\omega_E-\omega_N) + \\ +J(J_{ex}-\omega_E+\omega_N) + 6J(J_{ex}+\omega_N) + 6J(J_{ex}+\omega_E+\omega_N)\end{bmatrix} +$$

$$+ n\left(\begin{array}{c}\dfrac{\aleph_{(\text{HFC},\Sigma\text{HF}),\text{intra}}}{90}J(0) + \dfrac{\aleph_{(\text{HFC},\Sigma\text{HF}),\text{intra}}}{120}J(\omega_N) + \dfrac{\aleph_{(\text{HFC},\Sigma\text{HF}),\text{intra}}}{36}J(\omega_E) + \\ + \dfrac{\aleph_{(\text{HFC},\Delta\text{HF}),\text{intra}}}{720}\begin{bmatrix}3J(J_{ex}-\omega_E) + 3J(J_{ex}+\omega_E) + 12J(J_{ex}-\omega_E-\omega_N) + 12J(J_{ex}+\omega_E-\omega_N) + \\ +J(J_{ex}-\omega_E+\omega_N) + 6J(J_{ex}+\omega_N) + 6J(J_{ex}+\omega_E+\omega_N)\end{bmatrix}\end{array}\right) +$$

$$+ \frac{\left[4\aleph_{G,\Delta G} + 2\aleph_{EE,\Delta G}\right]}{240}J(J_{ex}-\omega_E) + \frac{\left[4\aleph_{G,\Delta G} + 2\aleph_{EE,-\Delta G}\right]}{240}J(J_{ex}+\omega_E) +$$

$$+ \frac{\left[9\Delta_{\text{EE}}^2 + 6\aleph_{G,\Sigma G}\right]}{180}J(\omega_E) + \frac{\Delta_{\text{EE}}^2}{30}J(2\omega_E)$$

(S4)

where the symbols are described in the Supplementary Information 5, and we have made a distinction between hyperfine couplings with n protons within the biradical ("intra"), and with one proton that belongs to the solvent ("inter") and could be a target of Overhauser DNP (ODNP). At the fields of interest in our studies, all spectral densities that have the electron and the nuclear Larmor frequency in them can be disregarded. This leaves solely the terms involving zero-frequency spectral densities:

$$R_{1,\text{BRW}} = -R\left[\hat{E}_Z\right] \simeq \frac{\Delta_{\text{EE}}^2}{90}J(0)$$

(S5)

and

$$R_{2,\text{BRW}} = -R\left[\hat{E}_+\right] = \frac{\Delta_{\text{EE}}^2}{36}J(0) + \frac{4\aleph_{G,\Sigma G}}{90}J(0) + \frac{\aleph_{(\text{HFC},\Sigma\text{HF}),\text{inter}}}{90}J(0) + n\left(\frac{\aleph_{(\text{HFC},\Sigma\text{HF}),\text{intra}}}{90}J(0)\right)$$

(S6)

$R_{1,\text{BRW}}$ is thus dominated by $\Delta_{\text{EE}}^2$, i.e., by the dipolar interaction between electrons; and $R_{2,\text{BRW}}$ is dominated by this term as well as by the second rank scalar products $\aleph_{G,\Sigma G}$ and $\aleph_{\text{HFC},\Sigma\text{HF}}$ between the *g*- tensor of one electron and the sum of the two $g$-tensors, and by the hyperfine coupling between one electron and one proton, and the sum of the hyperfines of the two electrons and one proton, respectively. The magnitudes of the coefficients in Eq. (S3) and Eq. (S4) for the case of a bistrityl-type biradical, are shown in the Table S2. These coefficients will scale according to spectral densities; Figure S1 show how these BRW-derived $R_1$ and $R_2$ terms change with the biradical's correlation time $\tau_c$. These rates should be compared with the relaxivity contributions arising from the vibrational local modes, which are likely to be in the order of what they were for the monotrityl radical i.e., $\approx 5 \cdot 10^4$ Hz. It follows that for the trityl bis-radical the contribution of these $R_{1,local}$ will no longer be dominant, primarily due to the onset of relaxation effects driven by $\Delta^2_{EE}$ and $\aleph_{G,\Sigma G}$ (the latter mostly at the high fields of interest; Table 1).

**Table S2:** Magnitude of the second-rank norm squared and the scalar products (defined in Supporting Information 5) predicted by Eqs. (S3) and (S6), for a trityl-based biradical with parameters as given in the Table 1.

| $\Delta^2_A$ / $\aleph_{A,B}$ | Magnitude / (rad/s)$^2$ |
|---|---|
| $\Delta^2_{EE}$ | $2 \times 10^{16}$ |
| $\aleph_{G,\Delta G}$ | 472 |
| $\aleph_{G,\Sigma G}$ | $2 \times 10^{17}$ |
| $\aleph_{EE,\Delta G}$ ($\aleph_{EE,-\Delta G}$) | 687699 / -687699 |
| $\aleph_{(HFC,\Sigma HF),intra}$/$\aleph_{(HFC,\Sigma HF),inter}$[1] | $1.30 \times 10^{13}$ / $4.9 \times 10^{13}$ |
| $\aleph_{(HFC,\Delta HF),intra}$/$\aleph_{(HFC,\Delta HF),inter}$[1] | $1.34 \times 10^{13}$ / $5.1 \times 10^{13}$ |

[1] The intra-molecular dipolar interaction between the electron and the methyl protons in each trityl group of the biradical was computed setting a proton-electron distance equal to 5.3 Å (between the proton and its closest electron; electron/proton hyperfine couplings between the two trityls in the molecule were disregarded). The inter-molecular dipolar interaction between the electron and the solvent was computed setting a proton-electron distance equal to 8.3 Å (again: between the proton and its closest electron only).

Rates do not change also with the increase of the intra-molecular protons n in Eqs. (S3) - (S6), since the term containing $\aleph_{(HFC,\Sigma HF),intra}$ remains much smaller than the term containing $\Delta^2_{EE}$, Fig. S1.

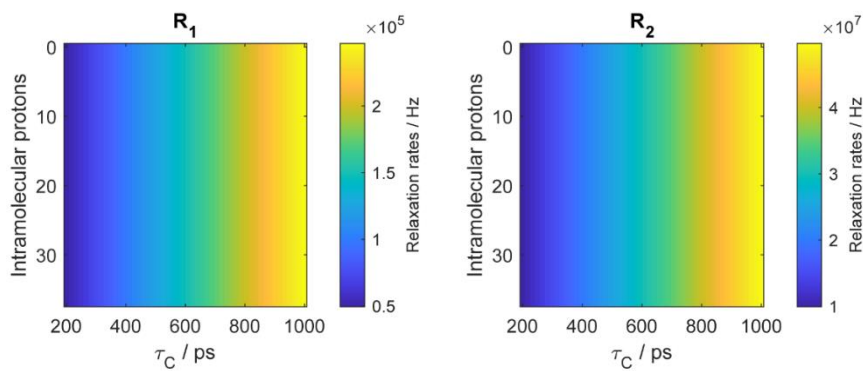

**Fig. S1:** $R_1$ and $R_2$ electron relaxation rates predicted by Eq. (S5) and Eq. (S6) as a function of the $\tau_C$ of the biradical/proton triad and on the number of closest intramolecular protons n to an electron (36 in the case of a model bis-trityl). For these calculations $J_{ex}$ was set equal to $+(\omega_E + \omega_N)$, $B_0$ was set equal to 14.08 T, and other simulation parameters are as given in the Table 1. The absence of a strong dependence with the number of protons reflects the dominant effects of electron-electron dipole and $g$-tensor anisotropies on the relaxation processes, over hyperfine counterparts.

It is enlightening to consider how this model predicts the electron saturation and J-DNP enhancement to proceed, as a function of the available microwave power. Considering that the rotational correlation time of the trityl monomer is about 150 ps –calculated from $\tau_C \simeq r_D^2/D_t$,[31] where $D_t$ is trityl's translational diffusion constant (in the order of $\sim 10^{-9}$ m$^2$ s$^{-1}$) and $r_D$ is the minimum distance approach between an electron in the biradical and a proton in the solvent (estimated at about 5-10 Å)– we decided to set the rotational correlation time for the biradical/proton triad equal to $\tau_C \approx 500$ ps (close to what's expected for a trityl biradical). Figure S2 shows the predicted electron saturation and J-DNP enhancement effects expected for this $\tau_C$ as a function of microwave off-resonance offset and nutation field; similar enhancements and saturation behaviours are predicted for smaller and higher rotational correlation times, up to 1 ns.

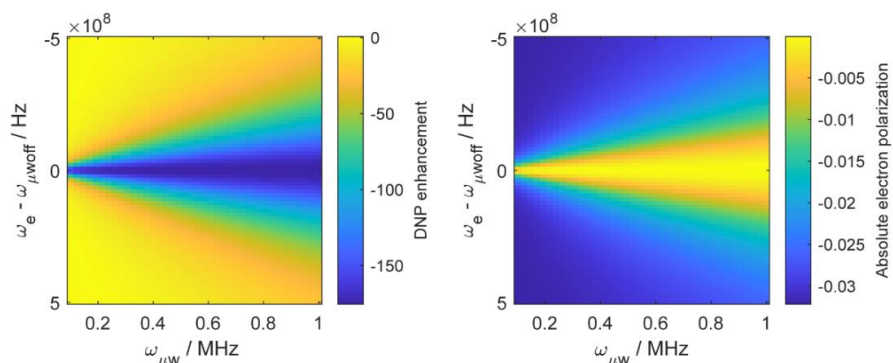

**Fig. S2:** Maximum enhancement (absolute value over the thermal equilibrium) in J-DNP (on the left), achieved between 80-100 ms (depending on the microwave power and Δω), and absolute electron polarization, $\hat{E}_Z$, (on the right), as a function of μw frequency offset from the free electron Larmor frequency and of the μw nutation power, $\omega_{\mu w}$. For these calculations $J_{ex}$ is equal to $\omega_E + \omega_N$, $B_0$ is 14.08 T, $\tau_C$ = 500 ps for the biradical/proton triad, and other simulation parameters as given in the Table 1.

A comparison between the J-DNP and the Overhauser DNP shows that in the electron/nucleus pair system, the ODNP case, the enhancement is strong if $B_0$ < 0.5T, but decays to negligible values if $B_0$ ≥ 3.4 T; this is as expected from classical theories [12,13,15,16,47], while the transient J-DNP is observed also at $B_0$ ≤ 3.4 T.

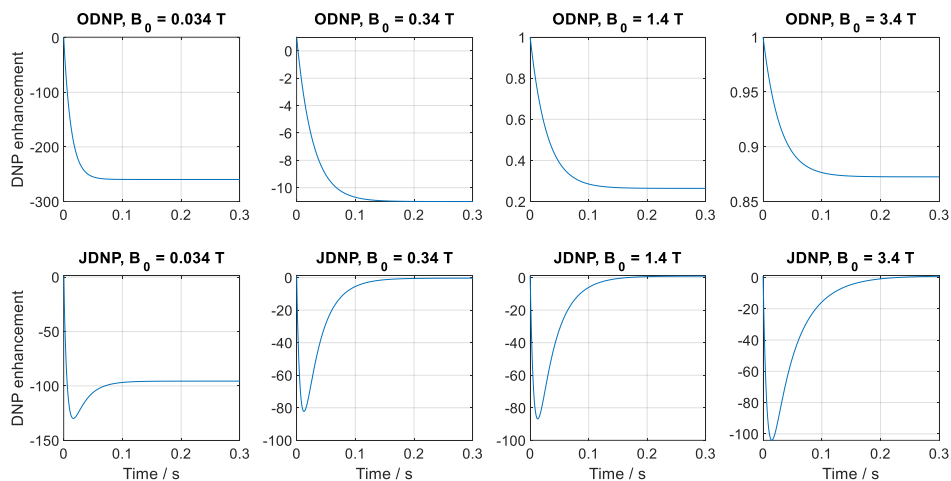

**Fig. S3:** Time domain simulations showing the evolution of the enhancement observed under continuous microwave irradiation of the electrons as a function of $B_0$, in ODNP (on the top) and in JDNP (on the bottom). For ODNP an electron/proton dipolar-coupled pair system with $\tau_c$ = 157 ps (typical of trityl [31]) was assumed; for J-DNP a biradical/proton dipole-coupled triad with $\tau_C$ = 500 ps for the for the biradical/proton triad and $J_{ex}$ = $+(\omega_E + \omega_N)$ was assumed. Other simulation parameters are given in the Table 1.

Overhauser DNP enhancements higher than those shown in Fig. S3, were observed at the magnetic fields of 1.4 T and 3.4 T, in water doped with trityl based mono radicals.[32,33] Rotational diffusion was the sole DNP-enabling contribution that was considered in our simulations, but translational diffusion can also become an important mechanism of polarization enhancement in actual solvents[32]– explaining the higher enhancements observed in the experimental measurements at these medium magnetic field strengths.

**Supporting Information 2: The Hamiltonian used in the propagation of the biradical/nuclear system – from the Cartesian to the triplet-singlet representations**

The laboratory frame Hamiltonian for a three-spin system composed by two electrons and one proton, where the electrons are connected by dipolar and exchange couplings, and the proton interacts with the electrons through hyperfine (dipolar) coupling only, can be written using single-spin Cartesian operators, as:

$$\hat{H}_{lab.} = \underbrace{\sum_k \hat{\vec{E}}^{(k)} \cdot \mathbf{Z}_E^{(k)} \cdot \vec{B}_0 + \hat{\vec{N}} \cdot \mathbf{Z}_N \cdot \vec{B}_0}_{\text{Zeeman interaction}} + \underbrace{\hat{\vec{E}}^{(1)} \cdot (\mathbf{D} + J_{ex}) \cdot \hat{\vec{E}}^{(2)}}_{\substack{\text{Dipolar and scalar} \\ \text{inter-electron interaction}}} + \underbrace{\sum_k \hat{\vec{E}}^{(k)} \cdot \mathbf{A}^{(k)} \cdot \hat{\vec{N}}}_{\text{Hyperfine interaction}} \quad (S7)$$

where $\mathbf{Z}_E^{(k)}$ are the Zeeman tensors (including g-anisotropy) of electron spins $k=1,2$; $\mathbf{Z}_N$ is the Zeeman tensor of the sole nuclear spin being considered (including the chemical shift anisotropy); $\mathbf{D}$ is the inter-electron dipolar interaction tensor in the point magnetic dipole approximation; $J_{ex}$ is the inter-electron scalar (*aka* "exchange") coupling in angular frequency units; $\mathbf{A}^{(k)}$ is the hyperfine interaction tensors of the nucleus with the indicated electron $k$; $\vec{B}_0$ is the external magnetic field; and $\hat{\vec{E}}^{(1)}$, $\hat{\vec{E}}^{(2)}$ and $\hat{\vec{N}}$ are the Cartesian spin-1/2 operators for the two electrons and nucleus. A rotating frame transformation with respect to the microwave frequency offset using the operator $\omega_{\mu woff} \sum_k \hat{E}_Z^{(k)}$ and preservation of the usual secular and pseudosecular terms, leads to:

$$\hat{H}_{rot.} = \omega_{\Sigma e}\left(\hat{E}_{1Z} + \hat{E}_{2Z}\right) + \omega_{\Delta e}\left(\hat{E}_{1Z} - \hat{E}_{2Z}\right) + \omega_N \hat{N}_Z + \omega_1 \hat{E}_{1Z}\hat{E}_{2Z} + \omega_2 \left(\frac{\hat{E}_{1+}\hat{E}_{2-} + \hat{E}_{1-}\hat{E}_{2+}}{2}\right)$$
$$+ A_\Sigma(\hat{E}_{1Z} + \hat{E}_{2Z})\hat{N}_Z + A_\Delta(\hat{E}_{1Z} - \hat{E}_{2Z})\hat{N}_Z + B_\Sigma(\hat{E}_{1Z} + \hat{E}_{2Z})\hat{N}_X + B_\Delta(\hat{E}_{1Z} - \hat{E}_{2Z})\hat{N}_X \quad (S8)$$

where $\omega_{\Sigma e}$ and $\omega_{\Delta e}$ are the sum and the difference between the rotating-frame offsets of the two electrons; $A_\Sigma = A_1 + A_2$ and $A_\Delta = A_1 - A_2$; $B_\Sigma = B_1 + B_2$ and $B_\Delta = B_1 - B_2$ are the sum and the difference of the secular and pseudo-secular coefficients describing the hyperfine interaction, respectively; and $\omega_1 = (J_{ex} + 2\mathbf{D})$ and $\omega_2 = (J_{ex} - \mathbf{D})$, where $\mathbf{D}$ is the inter-electron dipolar interaction tensor.

In the $J_{ex} \gg \omega_{\Delta e}$ case, where $\omega_{\Delta e} = \omega_{e1} - \omega_{e2}$ is the difference between the Larmor frequency of the two electrons, the electron Zeeman eigenstates $|\alpha_{e1}\beta_{e2}\rangle$ and $|\beta_{e1}\alpha_{e2}\rangle$ are no longer eigenfunctions of the spin Hamiltonian. We therefore express the Hamiltonian in the singlet/triplet electron basis sets [13,14]:

$$\left.\begin{array}{l}\left|\hat{T}_{+1}^{(e1,e2)}\right\rangle=\left|\alpha_{e1}\alpha_{e2}\right\rangle,\ \left|\hat{T}_{0}^{(e1,e2)}\right\rangle=\frac{1}{\sqrt{2}}\left(\left|\beta_{e1}\alpha_{e2}\right\rangle+\left|\alpha_{e1}\beta_{e2}\right\rangle\right),\\ \left|\hat{T}_{-1}^{(e1,e2)}\right\rangle=\left|\beta_{e1}\beta_{e2}\right\rangle,\ \left|\hat{S}_{0}^{(e1,e2)}\right\rangle=\frac{1}{\sqrt{2}}\left(\left|\beta_{e1}\alpha_{e2}\right\rangle-\left|\alpha_{e1}\beta_{e2}\right\rangle\right)\end{array}\right\}\otimes\{\alpha_{N},\beta_{N}\}=\left\{\begin{array}{l}\left|\hat{S}_{0}^{(e1,e2)},\alpha\right\rangle,\ \left|\hat{S}_{0}^{(e1,e2)},\beta\right\rangle\\ \left|\hat{T}_{0}^{(e1,e2)},\alpha\right\rangle,\ \left|\hat{T}_{0}^{(e1,e2)},\beta\right\rangle\\ \left|\hat{T}_{\pm1}^{(e1,e2)},\alpha\right\rangle,\ \left|\hat{T}_{\pm1}^{(e1,e2)},\beta\right\rangle\end{array}\right\} \quad (S9)$$

where the direct product indicates that the $\{|\alpha_N\rangle, |\beta_N\rangle\}$ Zeeman basis is preserved for the nucleus (proton). The transformation of the three-spin rotating frame Hamiltonian from a Cartesian basis to its singlet/triplet basis can be performed using fictitious spin-½ operators. [15,16] Since the addition of a third nuclear spin has to our knowledge not been reported before, we re-express the rotating-frame Hamiltonian in Eq. (S8), as a sum of terms containing direct products of operators acting in the $\hat{S}_0\hat{T}_0$ and the $\hat{T}_{+1}\hat{T}_{-1}$ subspaces, with the nuclear Zeeman base. The fictitious spin-1/2 operators that we use in this study to describe the $\hat{S}_0\hat{T}_0$ and the $\hat{T}_{+1}\hat{T}_{-1}$ subspaces follow from Vega's notation for two-spin fictitious operators[17] and were computed using the SpinDynamica software[18]:

$$\tilde{L}_{X}^{T_0S_0,\alpha} = \frac{1}{2}\left(\left|\hat{T}_0^{(e1,e2)},\alpha\right\rangle\left\langle\hat{S}_0^{(e1,e2)},\alpha\right|+\left|\hat{S}_0^{(e1,e2)},\alpha\right\rangle\left\langle\hat{T}_0^{(e1,e2)},\alpha\right|\right)$$
$$= \frac{\hat{E}_{1Z}}{4}-\frac{\hat{E}_{2Z}}{4}+\frac{1}{2}\left(\hat{E}_{1Z}\hat{N}_Z\right)-\frac{1}{2}\left(\hat{E}_{2Z}\hat{N}_Z\right) \quad (S10)$$

$$\tilde{L}_{X}^{T_0S_0,\beta} = \frac{1}{2}\left(\left|\hat{T}_0^{(e1,e2)},\beta\right\rangle\left\langle\hat{S}_0^{(e1,e2)},\beta\right|+\left|\hat{S}_0^{(e1,e2)},\beta\right\rangle\left\langle\hat{T}_0^{(e1,e2)},\beta\right|\right)$$
$$= \frac{\hat{E}_{1Z}}{4}-\frac{\hat{E}_{2Z}}{4}-\frac{1}{2}\left(\hat{E}_{1Z}\hat{N}_Z\right)+\frac{1}{2}\left(\hat{E}_{2Z}\hat{N}_Z\right) \quad (S11)$$

which we combine to obtain:

$$\omega_{\Delta e}\left(\hat{E}_{1Z}-\hat{E}_{2Z}\right)=2\omega_{\Delta e}\left(\tilde{L}_{X}^{T_0S_0,\alpha}+\tilde{L}_{X}^{T_0S_0,\beta}\right) \quad (S12)$$

$$A_{\Delta}\left(\hat{E}_{1Z}N_Z-\hat{E}_{2Z}N_Z\right)=A_{\Delta}\left(\tilde{L}_{X}^{T_0S_0,\alpha}-\tilde{L}_{X}^{T_0S_0,\beta}\right) \quad (S13)$$

The longitudinal fictitious ½-spin operators for the electron in the $\hat{S}_0\hat{T}_0$ are:

$$\tilde{L}_{Z}^{T_0S_0,\alpha} = \frac{1}{2}\left(\left|\hat{T}_0^{(e1,e2)},\alpha\right\rangle\left\langle\hat{T}_0^{(e1,e2)},\alpha\right|-\left|\hat{S}_0^{(e1,e2)},\alpha\right\rangle\left\langle\hat{S}_0^{(e1,e2)},\alpha\right|\right)$$
$$= -\frac{1}{2}\hat{E}_{1X}\hat{E}_{2X}-\frac{1}{2}\hat{E}_{1Y}\hat{E}_{2Y}-\hat{E}_{1X}\hat{E}_{2X}\hat{N}_Z-\hat{E}_{1Y}\hat{E}_{2Y}\hat{N}_Z \quad (S14)$$

$$\tilde{L}_{Z}^{T_0S_0,\beta} = \frac{1}{2}\left(\left|\hat{T}_0^{(e1,e2)},\beta\right\rangle\left\langle\hat{T}_0^{(e1,e2)},\beta\right|-\left|\hat{S}_0^{(e1,e2)},\beta\right\rangle\left\langle\hat{S}_0^{(e1,e2)},\beta\right|\right)$$
$$= -\frac{1}{2}\hat{E}_{1X}\hat{E}_{2X}-\frac{1}{2}\hat{E}_{1Y}\hat{E}_{2Y}+\hat{E}_{1X}\hat{E}_{2X}\hat{N}_Z+\hat{E}_{1Y}\hat{E}_{2Y}\hat{N}_Z \quad (S15)$$

$$\hat{E}^{T_0S_0,\alpha} = \left(\left|\hat{T}_0^{(e1,e2)},\alpha\right\rangle\left\langle\hat{T}_0^{(e1,e2)},\alpha\right|+\left|\hat{S}_0^{(e1,e2)},\alpha\right\rangle\left\langle\hat{S}_0^{(e1,e2)},\alpha\right|\right)$$
$$= -\hat{E}_{1Z}\hat{E}_{2Z}-2\hat{E}_{1Z}\hat{E}_{2Z}\hat{N}_Z+\frac{1}{2}\hat{N}_Z+\frac{\mathbf{E}}{4} \quad (S16)$$

$$\hat{E}^{T_0S_0,\beta} = \left(\left|\hat{T}_0^{(e1,e2)},\beta\right\rangle\left\langle\hat{T}_0^{(e1,e2)},\beta\right| + \left|\hat{S}_0^{(e1,e2)},\beta\right\rangle\left\langle\hat{S}_0^{(e1,e2)},\beta\right|\right)$$
$$= -\hat{E}_{1Z}\hat{E}_{2Z} + 2\hat{E}_{1Z}\hat{E}_{2Z}\hat{N}_Z - \frac{1}{2}\hat{N}_Z + \frac{\mathbf{E}}{4}$$
(S17)

from which we obtain:

$$\omega_2\left(\hat{L}_Z^{T_0S_0,\alpha} + \hat{L}_Z^{T_0S_0,\beta}\right) = \omega_2\left(\hat{E}_{1X}\hat{E}_{2X} + \hat{E}_{1Y}\hat{E}_{2Y}\right)$$
(S18)

$$\omega_1\left(\frac{\hat{E}^{T_0S_0,\alpha} + \hat{E}^{T_0S_0,\beta}}{4}\right) = \omega_1\left(-\frac{1}{2}\hat{E}_{1Z}\hat{E}_{2Z} + \frac{\mathbf{E}}{8}\right)$$
(S19)

The transverse fictitious ½-spin operators for the proton in the $\hat{S}_0\hat{T}_0$ subspace are:

$$\hat{L}_X^{S_0T_0,\alpha\beta} = \frac{1}{2}\left(\left|\hat{S}_0^{(e1,e2)},\alpha\right\rangle\left\langle\hat{T}_0^{(e1,e2)},\beta\right| + \left|\hat{T}_0^{(e1,e2)},\beta\right\rangle\left\langle\hat{S}_0^{(e1,e2)},\alpha\right|\right)$$
$$= \frac{1}{2}\hat{E}_{1Z}\hat{N}_X - \frac{1}{2}\hat{E}_{2Z}\hat{N}_X + \hat{E}_{1X}\hat{E}_{2Y}\hat{N}_Y - \hat{E}_{1Y}\hat{E}_{2X}\hat{N}_Y$$
(S20)

$$\hat{L}_X^{T_0S_0,\alpha\beta} = \frac{1}{2}\left(\left|\hat{T}_0^{(e1,e2)},\alpha\right\rangle\left\langle\hat{S}_0^{(e1,e2)},\beta\right| + \left|\hat{S}_0^{(e1,e2)},\beta\right\rangle\left\langle\hat{T}_0^{(e1,e2)},\alpha\right|\right)$$
$$= \frac{1}{2}\hat{E}_{1Z}\hat{N}_X - \frac{1}{2}\hat{E}_{2Z}\hat{N}_X - \hat{E}_{1X}\hat{E}_{2Y}\hat{N}_Y + \hat{E}_{1Y}\hat{E}_{2X}\hat{N}_Y$$
(S21)

from which we obtain the pseudo-secular component:

$$B_\Delta\left(\hat{L}_X^{S_0T_0,\alpha\beta} + \hat{L}_X^{T_0S_0,\alpha\beta}\right) = B_\Delta\left(\hat{E}_{1Z} - \hat{E}_{2Z}\right)\hat{N}_X$$
(S22)

Longitudinal fictitious ½-spin operators for the electron in the $\hat{T}_{+1}\hat{T}_{-1}$ subspace are:

$$\hat{L}_Z^{T_+T_{-1},\alpha} = \frac{1}{2}\left(\left|\hat{T}_{+1}^{(e1,e2)},\alpha\right\rangle\left\langle\hat{T}_{+1}^{(e1,e2)},\alpha\right| - \left|\hat{T}_{-1}^{(e1,e2)},\alpha\right\rangle\left\langle\hat{T}_{-1}^{(e1,e2)},\alpha\right|\right)$$
$$= \frac{\hat{E}_{1Z}}{4} + \frac{\hat{E}_{2Z}}{4} + \frac{1}{2}\left(\hat{E}_{1Z}\hat{N}_Z\right) + \frac{1}{2}\left(\hat{E}_{2Z}\hat{N}_Z\right)$$
(S23)

$$\hat{L}_Z^{T_+T_{-1},\beta} = \frac{1}{2}\left(\left|\hat{T}_{+1}^{(e1,e2)},\beta\right\rangle\left\langle\hat{T}_{+1}^{(e1,e2)},\beta\right| - \left|\hat{T}_{-1}^{(e1,e2)},\beta\right\rangle\left\langle\hat{T}_{-1}^{(e1,e2)},\beta\right|\right)$$
$$= \frac{\hat{E}_{1Z}}{4} + \frac{\hat{E}_{2Z}}{4} - \frac{1}{2}\left(\hat{E}_{1Z}\hat{N}_Z\right) - \frac{1}{2}\left(\hat{E}_{2Z}\hat{N}_Z\right)$$
(S24)

$$\hat{E}^{T_+T_{-1},\alpha} = \left(\left|\hat{T}_{+1}^{(e1,e2)},\alpha\right\rangle\left\langle\hat{T}_{+1}^{(e1,e2)},\alpha\right| + \left|\hat{T}_{-1}^{(e1,e2)},\alpha\right\rangle\left\langle\hat{T}_{-1}^{(e1,e2)},\alpha\right|\right)$$
$$= \hat{E}_{1Z}\hat{E}_{2Z} + 2\hat{E}_{1Z}\hat{E}_{2Z}\hat{N}_Z + \frac{\hat{N}_Z}{2} + \frac{\mathbf{E}}{4}$$
(S25)

$$\hat{E}^{T_+T_{-1},\beta} = \left(\left|\hat{T}_{+1}^{(e1,e2)},\beta\right\rangle\left\langle\hat{T}_{+1}^{(e1,e2)},\beta\right| + \left|\hat{T}_{-1}^{(e1,e2)},\beta\right\rangle\left\langle\hat{T}_{-1}^{(e1,e2)},\beta\right|\right)$$
$$= \hat{E}_{1Z}\hat{E}_{2Z} - 2\hat{E}_{1Z}\hat{E}_{2Z}\hat{N}_Z - \frac{\hat{N}_Z}{2} + \frac{\mathbf{E}}{4}$$
(S26)

from which we obtain:

$$\omega_{\Sigma e}\left(\hat{E}_{1Z} + \hat{E}_{2Z}\right) = 2\omega_{\Sigma e}\left(\tilde{L}_Z^{T_+T_{-1,\alpha}} + \tilde{L}_Z^{T_+T_{-1,\beta}}\right) \tag{S27}$$

$$A_{\Sigma}\left(\hat{E}_{1Z}\hat{N}_Z + \hat{E}_{2Z}\hat{N}_Z\right) = A_{\Sigma}\left(\tilde{L}_Z^{T_+T_{-1,\alpha}} - \tilde{L}_Z^{T_+T_{-1,\beta}}\right) \tag{S28}$$

$$\omega_1\left(\frac{1}{2}\hat{E}_{1Z}\hat{E}_{2Z} + \frac{\mathbf{E}}{8}\right) = \omega_1\left(\frac{\hat{E}^{T_+T_{-1,\alpha}} + \hat{E}^{T_+T_{-1,\beta}}}{4}\right) \tag{S29}$$

Finally, we introduce new transverse fictitious ½-spin operators for the proton in the $\hat{T}_{+1}\hat{T}_{-1}$ subspace given by:

$$\begin{aligned}\tilde{L}_X^{T_{+1,\alpha\beta}} &= \frac{1}{2}\left(\left|\hat{T}_{+1}^{(e1,e2)},\alpha\right\rangle\left\langle\hat{T}_{+1}^{(e1,e2)},\beta\right| + \left|\hat{T}_{+1}^{(e1,e2)},\beta\right\rangle\left\langle\hat{T}_{+1}^{(e1,e2)},\alpha\right|\right) \\ &= \frac{1}{2}\hat{E}_{1Z}\hat{N}_X + \frac{1}{2}\hat{E}_{2Z}\hat{N}_X + \hat{E}_{1Z}\hat{E}_{2Z}\hat{N}_X + \frac{\hat{N}_X}{4}\end{aligned} \tag{S30}$$

$$\begin{aligned}\tilde{L}_X^{T_{-1,\alpha\beta}} &= \frac{1}{2}\left(\left|\hat{T}_{-1}^{(e1,e2)},\alpha\right\rangle\left\langle\hat{T}_{-1}^{(e1,e2)},\beta\right| + \left|\hat{T}_{-1}^{(e1,e2)},\beta\right\rangle\left\langle\hat{T}_{-1}^{(e1,e2)},\alpha\right|\right) \\ &= -\frac{1}{2}\hat{E}_{1Z}\hat{N}_X - \frac{1}{2}\hat{E}_{2Z}\hat{N}_X + \hat{E}_{1Z}\hat{E}_{2Z}\hat{N}_X + \frac{\hat{N}_X}{4}\end{aligned} \tag{S31}$$

which are needed to describe the pseudo-secular component:

$$B_{\Sigma}\left(\tilde{L}_X^{T_{+1,\alpha\beta}} - \tilde{L}_X^{T_{-1,\alpha\beta}}\right) = B_{\Sigma}\left(\hat{E}_{1Z} + \hat{E}_{2Z}\right)\hat{N}_X \tag{S32}$$

The sum of these various terms, enables us to rewrite the Hamiltonian in Eq. (S8) under the action of microwave irradiation, as:

$$\hat{H}_{rot} = \hat{H}^{S_0T_0} + \hat{H}^{T_{+1}T_{-1}} + \hat{H}_{\mu w} \tag{S33}$$

where the term acting in the $\hat{S}_0\hat{T}_0$ space is

$$\begin{aligned}\hat{H}^{S_0T_0} &= 2\omega_{\Delta e}\left(\tilde{L}_X^{T_0S_{0,\alpha}} + \tilde{L}_X^{T_0S_{0,\beta}}\right) + A_{\Delta}\left(\tilde{L}_X^{T_0S_{0,\alpha}} - \tilde{L}_X^{T_0S_{0,\beta}}\right) + B_{\Delta}\left(\tilde{L}_X^{S_0T_{0,\alpha\beta}} + \tilde{L}_X^{T_0S_{0,\alpha\beta}}\right) + \\ &\quad -\omega_2\left(\tilde{L}_Z^{T_0S_{0,\alpha}} + \tilde{L}_Z^{T_0S_{0,\beta}}\right) - \omega_1\left(\frac{\hat{E}^{T_0S_{0,\alpha}} + \hat{E}^{T_0S_{0,\beta}}}{4}\right)\end{aligned} \tag{S34}$$

the term acting in the $\hat{T}_{+1}\hat{T}_{-1}$ space is:

$$\begin{aligned}\hat{H}^{T_{+1}T_{-1}} &= 2\omega_e\left(\tilde{L}_Z^{T_{\pm 1,\alpha}} + \tilde{L}_Z^{T_{\pm 1,\beta}}\right)\hat{R}_y^{T_{\pm 1}}(\theta) + \omega_N\hat{N}_Z + B_{\Sigma}\left(\tilde{L}_X^{T_{+1,\alpha\beta}} - \tilde{L}_X^{T_{-1,\alpha\beta}}\right) \\ &\quad + A_{\Sigma}\left(\tilde{L}_Z^{T_+T_{-1,\alpha}} - \tilde{L}_Z^{T_+T_{-1,\beta}}\right) + \omega_1\left(\frac{\hat{E}^{T_+T_{-1,\alpha}} + \hat{E}^{T_+T_{-1,\beta}}}{4}\right)\end{aligned} \tag{S35}$$

and $\hat{H}_{\mu w} = \omega_{\mu w}(\hat{E}_{1X} + \hat{E}_{2X})$ is the microwave irradiation operator, with $\omega_{\mu w}$ the strength of its nutation frequency. Here $\omega_{\Sigma e}$ and $\omega_{\Delta e} \sim 0$ are the sum and difference of the electron Larmor frequencies in the rotating frame, $\hat{R}_y^{T_{\pm 1}}(\theta)$ in Eq. (S35) is a rotation matrix about the y-axis that acts on both the α and β space; its associated $2\omega_e\left(\tilde{L}_Z^{T_{\pm 1,\alpha}} + \tilde{L}_Z^{T_{\pm 1,\beta}}\right)\hat{R}_y^{T_{\pm 1}}(\theta)$ term then corresponds to $\omega_e\cos(\theta)\left(\tilde{L}_Z^{T_{\pm 1,\alpha}} + \tilde{L}_Z^{T_{\pm 1,\beta}}\right) +$ $\omega_e \sin(\theta)(\hat{E}_{1X} + \hat{E}_{2X})$, with $\theta = \arctan(\omega_{\mu w}/\omega_{\Sigma e})$ the angle felt by the electron's effective field, and $\omega_e = \sqrt{\omega_{\Sigma e}^2 + \omega_{\mu w}^2}$ the effective field's strength. The matrix representation of Eq. (S33) is:

|  | $\left\|\hat{S}_0^{(e1,e2)},\alpha\right\rangle$ | $\left\|\hat{T}_{+1}^{(e1,e2)},\alpha\right\rangle$ | $\left\|\hat{T}_0^{(e1,e2)},\alpha\right\rangle$ | $\left\|\hat{T}_{-1}^{(e1,e2)},\alpha\right\rangle$ | $\left\|\hat{S}_0^{(e1,e2)},\beta\right\rangle$ | $\left\|\hat{T}_{+1}^{(e1,e2)},\beta\right\rangle$ | $\left\|\hat{T}_0^{(e1,e2)},\beta\right\rangle$ | $\left\|\hat{T}_{-1}^{(e1,e2)},\beta\right\rangle$ |
|---|---|---|---|---|---|---|---|---|
| $\left\|\hat{S}_0^{(e1,e2)},\alpha\right\rangle$ | $-\frac{\omega_J}{4} - \frac{\omega_\Sigma}{2} + \frac{\omega_N}{2}$ | 0 | $\frac{A_\Delta}{2} + \omega_{\Delta e}$ | 0 | 0 | 0 | $\frac{B_\Delta}{2}$ | 0 |
| $\left\|\hat{T}_{+1}^{(e1,e2)},\alpha\right\rangle$ | 0 | $\frac{A_\Sigma}{2} + \frac{\omega_J}{4} + \omega_e \cos(\theta) + \frac{\omega_N}{2}$ | $\frac{\omega_e \sin(\theta)}{\sqrt{2}}$ | 0 | 0 | $\frac{B_\Sigma}{2}$ | 0 | 0 |
| $\left\|\hat{T}_0^{(e1,e2)},\alpha\right\rangle$ | $\frac{A_\Delta}{2} + \omega_{\Delta e}$ | $\frac{\omega_e \sin(\theta)}{\sqrt{2}}$ | $-\frac{\omega_J}{4} + \frac{\omega_\Sigma}{2} + \frac{\omega_N}{2}$ | $\frac{\omega_e \sin(\theta)}{\sqrt{2}}$ | $\frac{B_\Delta}{2}$ | 0 | 0 | 0 |
| $\left\|\hat{T}_{-1}^{(e1,e2)},\alpha\right\rangle$ | 0 | 0 | $\frac{\omega_e \sin(\theta)}{\sqrt{2}}$ | $-\frac{A_\Sigma}{2} + \frac{\omega_J}{4} - \omega_e \cos(\theta) + \frac{\omega_N}{2}$ | 0 | 0 | 0 | $-\frac{B_\Sigma}{2}$ |
| $\left\|\hat{S}_0^{(e1,e2)},\beta\right\rangle$ | 0 | 0 | $\frac{B_\Delta}{2}$ | 0 | $-\frac{\omega_J}{4} - \frac{\omega_\Sigma}{2} - \frac{\omega_N}{2}$ | 0 | $-\frac{A_\Delta}{2} + \omega_{\Delta e}$ | 0 |
| $\left\|\hat{T}_{+1}^{(e1,e2)},\beta\right\rangle$ | 0 | $\frac{B_\Sigma}{2}$ | 0 | 0 | 0 | $-\frac{A_\Sigma}{2} + \frac{\omega_J}{4} + \omega_e \cos(\theta) - \frac{\omega_N}{2}$ | $\frac{\omega_e \sin(\theta)}{\sqrt{2}}$ | 0 |
| $\left\|\hat{T}_0^{(e1,e2)},\beta\right\rangle$ | $\frac{B_\Delta}{2}$ | 0 | 0 | 0 | $-\frac{A_\Delta}{2} + \omega_{\Delta e}$ | $\frac{\omega_e \sin(\theta)}{\sqrt{2}}$ | $-\frac{\omega_J}{4} + \frac{\omega_\Sigma}{2} - \frac{\omega_N}{2}$ | $\frac{\omega_e \sin(\theta)}{\sqrt{2}}$ |
| $\left\|\hat{T}_{-1}^{(e1,e2)},\beta\right\rangle$ | 0 | 0 | 0 | $-\frac{B_\Sigma}{2}$ | 0 | 0 | $\frac{\omega_e \sin(\theta)}{\sqrt{2}}$ | $\frac{A_\Sigma}{2} + \frac{\omega_J}{4} - \omega_e \cos(\theta) - \frac{\omega_N}{2}$ |

(S36)

Notice that these microwave-related terms act solely within the triplet manifold mixing the $\hat{T}_{\pm 1}^{(e1,e2)}$ and $\hat{T}_0^{(e1,e2)}$ states, but do not involve the $\hat{S}_0^{(e1,e2)}$ singlet. The latter, however, is not isolated: it gets connected to $\hat{T}_0^{(e1e2)}$ via the difference in secular hyperfine couplings with the nucleus $A_\Delta$. Additional simulations –not shown– demonstrate that the pseudosecular terms are not essential for describing the J-DNP effect.

**Supporting Information 3: Defining the biradical/nuclear system population operators describing the J-DNP enhancement**

The main text defines $\hat{S}_0 \hat{N}_Z$, $\hat{T}_{\pm 1} \hat{N}_Z$ and $\hat{T}_0 \hat{N}_Z$ operators and relates these to the differences between the population operators $O_\alpha$ and $O_\beta$ (defined below), leading to the predicted nuclear polarization enhancement. As these three-spin states have to our knowledge not been previously defined, we summarize them here. To do this we rely again on Vega's two-spin triplet/singlet (TS) population operators, [15] and we direct-product them with a nuclear spin state that can be in either the α or β state. These can be written in terms of single-spin product operators, as:

$$\hat{S}_{0,\alpha} = \left| \hat{S}_0^{(e1,e2)}, \alpha \right\rangle \left\langle \hat{S}_0^{(e1,e2)}, \alpha \right| = \frac{\mathbf{E}}{8} + \frac{\hat{N}_Z}{4} - \frac{1}{4}\left(\hat{E}_{1-}\hat{E}_{2+}\right) - \frac{1}{4}\left(\hat{E}_{1+}\hat{E}_{2-}\right) + $$
$$- \frac{1}{2}\left(\hat{E}_{1Z}\hat{E}_{2Z}\right) - \frac{1}{2}\left(\hat{E}_{1-}\hat{E}_{2+}\hat{N}_Z\right) - \frac{1}{2}\left(\hat{E}_{1+}\hat{E}_{2-}\hat{N}_Z\right) - \hat{E}_{1Z}\hat{E}_{2Z}\hat{N}_Z \quad (S37)$$

$$\hat{T}_{\pm 1,\alpha} = \left| \hat{T}_{\pm 1}^{(e1,e2)}, \alpha \right\rangle \left\langle \hat{T}_{\pm 1}^{(e1,e2)}, \alpha \right| = \frac{\mathbf{E}}{8} + \frac{\hat{N}_Z}{4} + \frac{\hat{E}_{1Z}}{4} \pm \frac{\hat{E}_{2Z}}{4} + $$
$$+ \frac{1}{2}\left(\hat{E}_{1Z}\hat{E}_{2Z}\right) \pm \frac{1}{2}\left(\hat{E}_{1Z}\hat{N}_Z\right) \pm \frac{1}{2}\left(\hat{E}_{2Z}\hat{N}_Z\right) + \hat{E}_{1Z}\hat{E}_{2Z}\hat{N}_Z \quad (S38)$$

$$\hat{T}_{0,\alpha} = \left| \hat{T}_0^{(e1,e2)}, \alpha \right\rangle \left\langle \hat{T}_0^{(e1,e2)}, \alpha \right| = \frac{\mathbf{E}}{8} + \frac{\hat{N}_Z}{4} + \frac{1}{4}\left(\hat{E}_{1-}\hat{E}_{2+}\right) + \frac{1}{4}\left(\hat{E}_{1+}\hat{E}_{2-}\right) + $$
$$- \frac{1}{2}\left(\hat{E}_{1Z}\hat{E}_{2Z}\right) + \frac{1}{2}\left(\hat{E}_{1-}\hat{E}_{2+}\hat{N}_Z\right) + \frac{1}{2}\left(\hat{E}_{1+}\hat{E}_{2-}\hat{N}_Z\right) - \hat{E}_{1Z}\hat{E}_{2Z}\hat{N}_Z \quad (S39)$$

$$\hat{S}_{0,\beta} = \left| \hat{S}_0^{(e1,e2)}, \beta \right\rangle \left\langle \hat{S}_0^{(e1,e2)}, \beta \right| = \frac{\mathbf{E}}{8} - \frac{\hat{N}_Z}{4} - \frac{1}{4}\left(\hat{E}_{1-}\hat{E}_{2+}\right) - \frac{1}{4}\left(\hat{E}_{1+}\hat{E}_{2-}\right) - $$
$$+ \frac{1}{2}\left(\hat{E}_{1Z}\hat{E}_{2Z}\right) + \frac{1}{2}\left(\hat{E}_{1-}\hat{E}_{2+}\hat{N}_Z\right) + \frac{1}{2}\left(\hat{E}_{1+}\hat{E}_{2-}\hat{N}_Z\right) + \hat{E}_{1Z}\hat{E}_{2Z}\hat{N}_Z \quad (S40)$$

$$\hat{T}_{\pm 1,\beta} = \left| \hat{T}_{\pm 1}^{(e1,e2)}, \beta \right\rangle \left\langle \hat{T}_{\pm 1}^{(e1,e2)}, \beta \right| = \frac{\mathbf{E}}{8} - \frac{\hat{N}_Z}{4} \pm \frac{\hat{E}_{1Z}}{4} \pm \frac{\hat{E}_{2Z}}{4} + $$
$$+ \frac{1}{2}\left(\hat{E}_{1Z}\hat{E}_{2Z}\right) \mp \frac{1}{2}\left(\hat{E}_{1Z}\hat{N}_Z\right) \mp \frac{1}{2}\left(\hat{E}_{2Z}\hat{N}_Z\right) - \hat{E}_{1Z}\hat{E}_{2Z}\hat{N}_Z \quad (S41)$$

$$\hat{T}_{0,\beta} = \left| \hat{T}_0^{(e1,e2)}, \beta \right\rangle \left\langle \hat{T}_0^{(e1,e2)}, \beta \right| = \frac{\mathbf{E}}{8} - \frac{\hat{N}_Z}{4} + \frac{1}{4}\left(\hat{E}_{1-}\hat{E}_{2+}\right) + \frac{1}{4}\left(\hat{E}_{1+}\hat{E}_{2-}\right) - $$
$$+ \frac{1}{2}\left(\hat{E}_{1Z}\hat{E}_{2Z}\right) - \frac{1}{2}\left(\hat{E}_{1-}\hat{E}_{2+}\hat{N}_Z\right) - \frac{1}{2}\left(\hat{E}_{1+}\hat{E}_{2-}\hat{N}_Z\right) + \hat{E}_{1Z}\hat{E}_{2Z}\hat{N}_Z \quad (S42)$$

Taking suitable differences among these states, leads to the longitudinal fictitious operators used in the main text (Figures 4-5) to describe how singlet and triplet states enhance the nuclear polarization:

$$\hat{S}_0^{(e1,e2)} \hat{N}_Z = \frac{1}{2}\left( \left| \hat{S}_0^{(e1,e2)}, \alpha \right\rangle \left\langle \hat{S}_0^{(e1,e2)}, \alpha \right| - \left| \hat{S}_0^{(e1,e2)}, \beta \right\rangle \left\langle \hat{S}_0^{(e1,e2)}, \beta \right| \right)$$
$$= \frac{\hat{N}_Z}{4} - \frac{1}{2}\left( \hat{E}_{1-}\hat{E}_{2+}\hat{N}_Z + \hat{E}_{1+}\hat{E}_{2-}\hat{N}_Z \right) - \hat{E}_{1Z}\hat{E}_{2Z}\hat{N}_Z \quad (S43)$$

$$\hat{T}_{\pm 1}^{(e1,e2)}\hat{N}_Z = \frac{1}{2}\left(\left|\hat{T}_{\pm 1}^{(e1,e2)},\alpha\right\rangle\left\langle\hat{T}_{\pm 1}^{(e1,e2)},\alpha\right| - \left|\hat{T}_{\pm 1}^{(e1,e2)},\beta\right\rangle\left\langle\hat{T}_{\pm 1}^{(e1,e2)},\beta\right|\right)$$

$$= \frac{\hat{N}_Z}{4} \pm \frac{1}{2}\left(\hat{E}_{1Z}\hat{N}_Z + \hat{E}_{2Z}\hat{N}_Z\right) + \hat{E}_{1Z}\hat{E}_{2Z}\hat{N}_Z \quad \text{(S44)}$$

$$\hat{T}_0^{(e1,e2)}\hat{N}_Z = \frac{1}{2}\left(\left|\hat{T}_0^{(e1,e2)},\alpha\right\rangle\left\langle\hat{T}_0^{(e1,e2)},\alpha\right| - \left|\hat{T}_0^{(e1,e2)},\beta\right\rangle\left\langle\hat{T}_0^{(e1,e2)},\beta\right|\right)$$

$$= \frac{\hat{N}_Z}{4} + \frac{1}{2}\left(\hat{E}_{1-}\hat{E}_{2+}\hat{N}_Z + \hat{E}_{1+}\hat{E}_{2-}\hat{N}_Z\right) - \hat{E}_{1Z}\hat{E}_{2Z}\hat{N}_Z \quad \text{(S45)}$$

**Supporting Information 4: Additional Redfield-derived relaxation rates for the biradical/nuclear system**

The main text presented a simplified version of the relaxation rates of $\hat{S}_{0,\alpha/\beta}$, $\hat{T}_{0,\alpha/\beta}$ and $\hat{T}_{\pm 1,\alpha,\beta}$, whose full expressions are provided here. For $\hat{S}_{0,\alpha/\beta}$ these were:

$$-R\left[\hat{S}_{0,\beta}\right] = \frac{\Delta_{\Delta HF}^2}{180} J(J_{ex} - \omega_E + \omega_N) + \frac{\Delta_{\Delta HF}^2}{30} J(J_{ex} + \omega_E + \omega_N) + \frac{\Delta_{CSA}^2}{15} J(\omega_N) + \frac{\Delta_{\Delta HF}^2}{60} J(J_{ex} + \omega_N) +$$
$$+ \frac{\left[\Delta_{\Delta HF}^2 + 4\aleph_{\Delta G,\Delta G\text{-}\Delta HF}\right]}{90} J(J_{ex}) + \frac{\left[\Delta_{\Delta HF}^2 + 4\aleph_{\Delta G,\Delta G\text{-}\Delta HF}\right]}{120} J(J_{ex} - \omega_E) + \frac{\left[\Delta_{\Delta HF}^2 + 4\aleph_{\Delta G,\Delta G\text{-}\Delta HF}\right]}{120} J(J_{ex} + \omega_E)$$
(S46)

and

$$-R\left[\hat{S}_{0,\alpha}\right] = \frac{\Delta_{\Delta HF}^2}{180} J(J_{ex} + \omega_E - \omega_N) + \frac{\Delta_{\Delta HF}^2}{30} J(J_{ex} - \omega_E - \omega_N) + \frac{\Delta_{CSA}^2}{15} J(\omega_N) + \frac{\Delta_{\Delta HF}^2}{60} J(J_{ex} - \omega_N) +$$
$$+ \frac{\left[\Delta_{\Delta HF}^2 + 4\aleph_{\Delta G,\Delta G+\Delta HF}\right]}{90} J(J_{ex}) + \frac{\left[\Delta_{\Delta HF}^2 + 4\aleph_{\Delta G,\Delta G+\Delta HF}\right]}{120} J(J_{ex} - \omega_E) + \frac{\left[\Delta_{\Delta HF}^2 + 4\aleph_{\Delta G,\Delta G+\Delta HF}\right]}{120} J(J_{ex} + \omega_E)$$
(S47)

For the $\hat{T}_{+1,\alpha/\beta}$ states these were:

$$-R\left[\hat{T}_{+1,\beta}\right] = \frac{\Delta_{\Delta HF}^2}{180} J(J_{ex} + \omega_E - \omega_N) + \frac{\left[\Delta_{\Delta HF}^2 + 4\aleph_{\Delta G,\Delta G\text{-}\Delta HF}\right]}{120} J(J_{ex} + \omega_E) +$$
$$+ \frac{\left[\Delta_{\Sigma HF}^2 + 4\aleph_{CSA,CSA+\Sigma HF}\right]}{60} J(\omega_N) + \frac{\left[5\Delta_{\Sigma HF}^2 + 12\aleph_{EE+\Sigma G,(EE+\Sigma G)\text{-}\Sigma HF}\right]}{360} J(\omega_E) + \frac{\Delta_{EE}^2}{15} J(2\omega_E)$$
(S48)

and

$$-R\left[\hat{T}_{+1,\alpha}\right] = \frac{\Delta_{\Delta HF}^2}{30} J(J_{ex} + \omega_E + \omega_N) + \frac{\left[\Delta_{\Delta HF}^2 + 4\aleph_{\Delta G,\Delta G+\Delta HF}\right]}{120} J(J_{ex} + \omega_E) +$$
$$+ \frac{\left[\Delta_{\Sigma HF}^2 + 4\aleph_{CSA,CSA+\Sigma HF}\right]}{60} J(\omega_N) + \frac{\left[5\Delta_{\Sigma HF}^2 + 4\aleph_{EE+\Sigma G,(EE+\Sigma G)+\Sigma HF}\right]}{120} J(\omega_E) + \frac{\Delta_{EE}^2}{15} J(2\omega_E)$$
(S49)

For the $\hat{T}_{0,\alpha/\beta}$, the self-relaxation rates are

$$-R\left[\hat{T}_{0,\beta}\right] = \frac{\Delta_{CSA}^2}{15} J(\omega_N) + \frac{\Delta_{\Delta HF}^2}{60} J(J_{ex} - \omega_N) +$$
$$+ \frac{\left[\Delta_{\Delta HF}^2 + 4\aleph_{\Delta G,\Delta G-\Delta HF}\right]}{90} J(J_{ex}) + \frac{\left[6\Delta_{EE}^2 + 5\Delta_{\Sigma HF}^2 + 6\aleph_{\Sigma G,\Sigma G-\Sigma HF}\right]}{90} J(\omega_E)$$
(S50)

and

$$-R\left[\hat{T}_{0,\alpha}\right] = \frac{\Delta_{CSA}^2}{15} J(\omega_N) + \frac{\Delta_{\Delta HF}^2}{60} J(J_{ex} + \omega_N) +$$
$$+ \frac{\left[\Delta_{\Delta HF}^2 + 4\aleph_{\Delta G,\Delta G+\Delta HF}\right]}{90} J(J_{ex}) + \frac{\left[6\Delta_{EE}^2 + 5\Delta_{\Sigma HF}^2 + 6\aleph_{\Sigma G,\Sigma G+\Sigma HF}\right]}{90} J(\omega_E)$$
(S51)

And for $\hat{T}_{-1,\alpha/\beta}$ the self-relaxation rates were:

$$-R\left[\hat{T}_{-1,\beta}\right] = \frac{\Delta_{\Delta HF}^2}{30} J(J_{ex} - \omega_E - \omega_N) + \frac{\left[\Delta_{\Delta HF}^2 + 4\aleph_{\Delta G,\Delta G-\Delta HF}\right]}{120} J(J_{ex} - \omega_E) +$$
$$+ \frac{\left[\Delta_{\Sigma HF}^2 + 4\aleph_{CSA,CSA-\Sigma HF}\right]}{60} J(\omega_N) + \frac{\left[5\Delta_{\Sigma HF}^2 + 4\aleph_{EE-\Sigma G,EE-\Sigma G+\Sigma HF}\right]}{120} J(\omega_E) + \frac{\Delta_{EE}^2}{15} J(2\omega_E)$$
(S52)

and

$$-R\left[\hat{T}_{-1,\alpha}\right] = \frac{\Delta_{\Delta HF}^2}{180} J\left(J_{ex} - \omega_E + \omega_N\right) + \frac{\left[\Delta_{\Delta HF}^2 + 4\aleph_{\Delta G, \Delta G + \Delta HF}\right]}{120} J\left(J_{ex} - \omega_E\right) +$$
$$+ \frac{\left[\Delta_{\Sigma HF}^2 + 4\aleph_{CSA, CSA-\Sigma HF}\right]}{60} J\left(\omega_N\right) + \frac{\left[5\Delta_{\Sigma HF}^2 + 12\aleph_{EE-\Sigma G, EE-\Sigma G-\Sigma HF}\right]}{360} J\left(\omega_E\right) + \frac{\Delta_{EE}^2}{15} J\left(2\omega_E\right)$$

(S53)

These expressions were all derived taking the possibility of having the spins' relaxation driven by the nuclear chemical shift anisotropy tensor (**CSA**), by $\Delta \mathbf{G} = \mathbf{G_1} - \mathbf{G_2}$ and by $\Delta \mathbf{HF} = \mathbf{HFC_1} - \mathbf{HFC_2}$ anisotropies deriving from tensors associated to the differences between the two *g*- and electron/nuclear hyperfine coupling tensors, respectively; by tensors $\mathbf{\Sigma G} = \mathbf{G_1} + \mathbf{G_2}$ and $\mathbf{\Sigma HF} = \mathbf{HFC_1} + \mathbf{HFC_2}$ associated to the sums of these two electron *g*- and hyperfine tensors, and by the **EE** interaction representing the dipolar tensor between the two. As is usual in spin relaxation theory [19,20], all these rates contain combinations of second-rank norms squared $\Delta_A^2$ of all the aforementioned tensors **A**, second-rank scalar products $\aleph_{A,B}$ of 3x3 tensors **A** and **B** [3], and linear combinations of these products among various tensors, as given in Supporting Information 5. Figures S4 and S5 below expand this matter further, by showing how rates of $\hat{S}_{0,\alpha/\beta}$, $\hat{T}_{0,\alpha/\beta}$ and $\hat{T}_{\pm 1,\alpha/\beta}$ vary, when $J_{ex}$ matches $\pm(\omega_E + \omega_N)$ – this time as a function of $B_0$ and $\tau_C$.

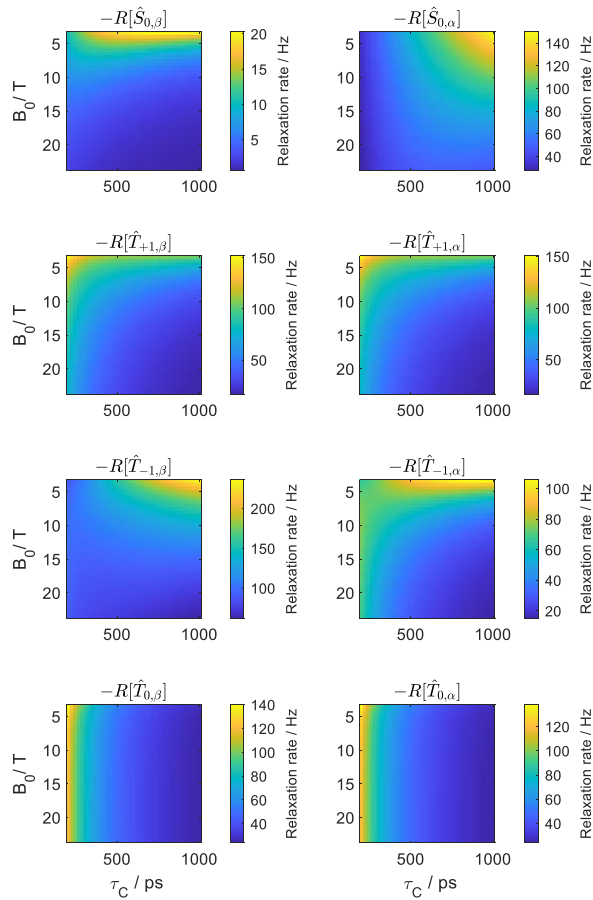

**Fig. S4:** Numerically calculated self-relaxation rates of $\hat{S}_{0,\alpha/\beta}$, $\hat{T}_{0,\alpha/\beta}$ and $\hat{T}_{\pm 1,\alpha/\beta}$ as a function of $B_0$ and of the $\tau_C$ of the biradical/proton triad, when $J_{ex}$ is positive and equal to $J_{ex} = -(\omega_E + \omega_N)$. Other simulation parameters are given in Table 1.

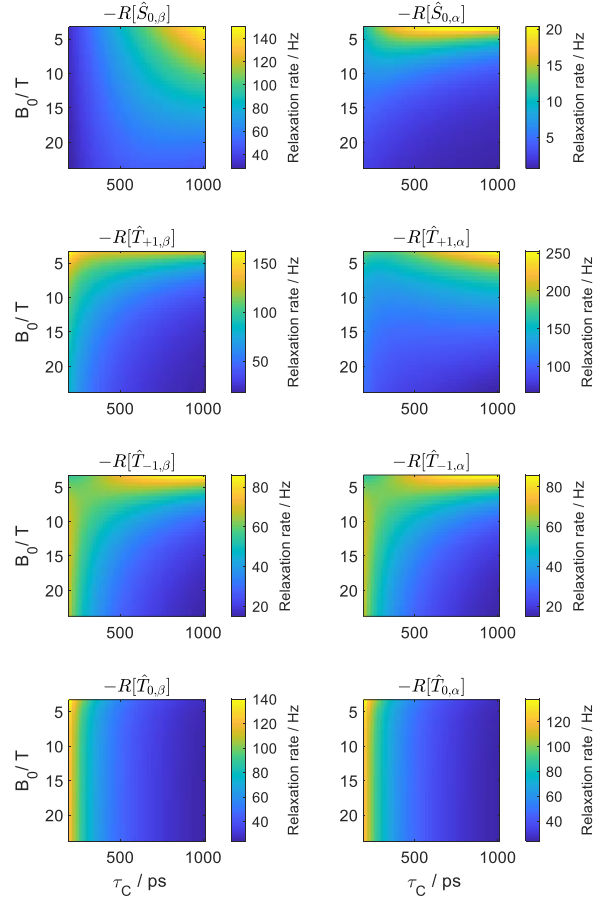

**Fig. S5:** Numerically calculated self-relaxation rates of $\hat{S}_{0,\alpha/\beta}$, $\hat{T}_{0,\alpha/\beta}$ and $\hat{T}_{\pm 1,\alpha/\beta}$ as a function of $B_0$ and of the $\tau_C$ of the biradical/proton triad, when $J_{ex}$ is negative and equal to $J_{ex} = +(\omega_E + \omega_N)$. Other simulation parameters are given in Table 1.

Notice how these rates decrease with magnetic field and change differentially for α/β states with the correlation time. The main text presented how the rates of $\hat{S}_{0,\alpha/\beta}$, $\hat{T}_{0,\alpha/\beta}$ and $\hat{T}_{\pm 1,\alpha/\beta}$ changed with exchange coupling and magnetic field, according to numerical and analytical predictions. For completion this section derives the expressions predicted by this theory, for the $\hat{N}_Z\hat{T}_{\pm 1}$, $\hat{N}_Z\hat{T}_0$ and $\hat{S}_0\hat{N}_Z$ states.

$$-R\left[\hat{N}_Z\hat{S}_0\right] = \frac{2\Delta_{CSA}^2}{15}J(\omega_N) + \frac{\Delta_{\Delta HF}^2}{120}\left[J(J_{ex} - \omega_N) + J(J_{ex} + \omega_N)\right] + $$
$$+ \frac{\Delta_{\Delta HF}^2}{60}\left[J(J_{ex} - \omega_E - \omega_N) + J(J_{ex} + \omega_E + \omega_N)\right] + \quad (S54)$$
$$+ \frac{\Delta_{\Delta HF}^2}{360}\left[J(J_{ex} + \omega_E - \omega_N) + J(J_{ex} - \omega_E + \omega_N)\right] + $$
$$+ \frac{\Delta_{\Delta HF}^2 + 4\Delta_{\Delta G}^2}{90}J(J_{ex}) + \frac{\Delta_{\Delta HF}^2 + 4\Delta_{\Delta G}^2}{120}\left[J(J_{ex} - \omega_E) + J(J_{ex} + \omega_E)\right]$$

$$-R\left[\hat{N}_Z\hat{T}_{+1}\right] = \frac{\Delta_{EE}^2}{15}J(2\omega_E) + \frac{\Delta_{\Delta HF}^2}{360}J(J_{ex} + \omega_E - \omega_N) + \frac{\Delta_{\Delta HF}^2}{60}J(J_{ex} + \omega_E + \omega_N) + $$
$$\frac{\Delta_{\Delta HF}^2 + 4\Delta_{\Delta G}^2}{120}J(J_{ex} + \omega_E) + \frac{\left[\Delta_{\Sigma HF}^2 + 4\aleph_{CSA,CSA+\Sigma HF}\right]}{30}J(\omega_N) + \frac{\left[5\Delta_{\Sigma HF}^2 + 6\Delta_{EE+\Sigma G}^2\right]}{180}J(\omega_E) \quad (S55)$$

$$-R\left[\hat{N}_Z\hat{T}_{-1}\right] = \frac{\Delta_{EE}^2}{15}J(2\omega_E) + \frac{\Delta_{\Delta HF}^2}{360}J(J_{ex}-\omega_E+\omega_N) + \frac{\Delta_{\Delta HF}^2}{60}J(J_{ex}-\omega_E-\omega_N) +$$
$$\frac{\Delta_{\Delta HF}^2 + 4\Delta_{\Delta G}^2}{120}J(J_{ex}-\omega_E) + \frac{\left[\Delta_{\Sigma HF}^2 + 4\aleph_{CSA,CSA-\Sigma HF}\right]}{30}J(\omega_N) + \frac{\left[5\Delta_{\Sigma HF}^2 + 6\Delta_{EE-\Sigma G}^2\right]}{180}J(\omega_E)$$
(S56)

$$-R\left[\hat{N}_Z\hat{T}_0\right] = \frac{2\Delta_{CSA}^2}{15}J(\omega_N) + \frac{\Delta_{\Delta HF}^2}{120}\left[J(J_{ex}-\omega_N) + J(J_{ex}+\omega_N)\right] +$$
$$+\frac{\Delta_{\Delta HF}^2 + 4\Delta_{\Delta G}^2}{90}J(J_{ex}) + \frac{\left[6\Delta_{EE}^2 + 5\Delta_{\Sigma HF}^2 + 6\Delta_{\Sigma G}^2\right]}{90}J(\omega_E)$$
(S57)

where the meaning of the various constants and functions are the same as in Eqs. (S46) - (S53). Figures S6 and S7 present how these rates depend on the magnetic fields and on the rotational correlation times.

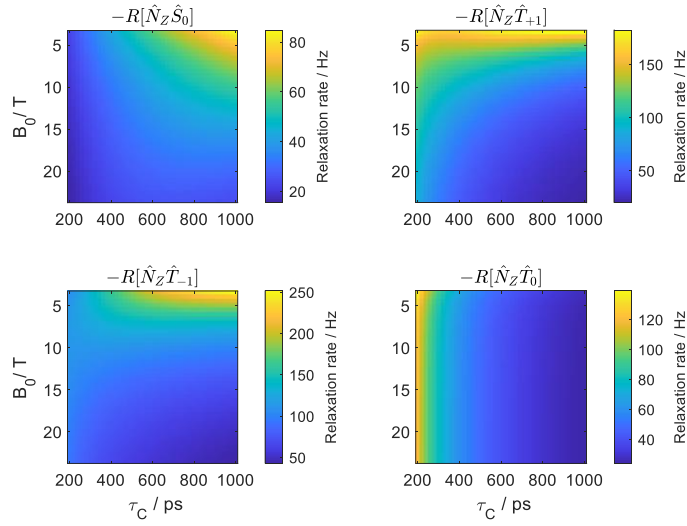

**Fig. S6:** Self-relaxation rates of $\hat{N}_Z\hat{S}_0$, $\hat{N}_Z\hat{T}_{\pm 1}$ and $\hat{N}_Z\hat{T}_0$ states calculated as a function of $B_0$ and of the $\tau_C$ of the biradical/proton triad, for $J_{ex} = -(\omega_E + \omega_N)$. Other simulation parameters are given in Table 1.

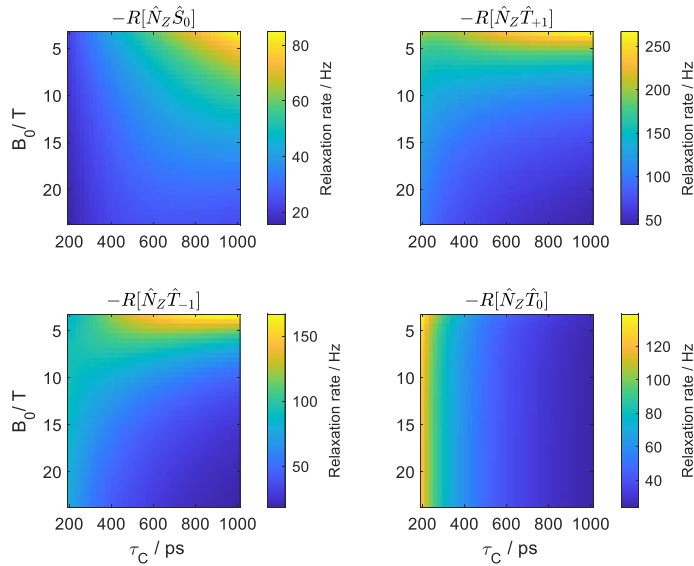

**Fig. S7:** Self-relaxation rates of $\hat{N}_Z\hat{S}_0$, $\hat{N}_Z\hat{T}_{\pm 1}$ and $\hat{N}_Z\hat{T}_0$ states calculated as a function of $B_0$ and of the $\tau_C$ of the biradical/proton triad, at $J_{ex} = +(\omega_E + \omega_N)$. Other simulation parameters are given in Table 1.

**Supporting Information 5: Additional information about the Redfield analysis of the biradical/nuclear system**

The rate expressions derived by the Redfield theory analysis for the three-spin system in Eqs. (5)-(12) of the main text, and in Eqs. S1 - S6 and S46 -S57 in the Supporting Information, were expressed on the basis of the second-rank square norms and scalar products of 3x3 tensors:

$$\aleph_{A,B} = \frac{\Delta^2_{A+B} - \Delta^2_{A-B}}{4}$$

$$\Delta^2_A = a^2_{XX} + a^2_{YY} + a^2_{ZZ} - a_{XX}a_{YY} - a_{XX}a_{ZZ} - a_{YY}a_{ZZ} + \qquad \text{(S58)}$$

$$+ \frac{3}{4}\left[(a_{XY} + a_{YX})^2 + (a_{XZ} + a_{ZX})^2 + (a_{YZ} + a_{ZY})^2\right]$$

where $\Delta_A$ is the second-rank norm squared of a tensor **A** and $\aleph_{A,B}$ is the second-rank scalar product between two 3×3 interaction tensors **A** and **B**. They can also contain linear combinations of more than two tensors and are often expressed based on the following algebraic relation:

$$\begin{aligned}
&\Delta^2_{A-B} = \Delta^2_A + \Delta^2_B - 2\aleph_{A,B}, \quad \Delta^2_{A+B} = \Delta^2_A + \Delta^2_B + 2\aleph_{A,B}, \\
&\Delta^2_{A+(B+C)} = \Delta^2_A + \Delta^2_{B+C} + 2\aleph_{A,B+C}, \quad \Delta^2_{A-(B+C)} = \Delta^2_A + \Delta^2_{B+C} - 2\aleph_{A,B+C}, \\
&\Delta^2_{A-B} = \aleph_{A-B,A-B}, \quad \Delta^2_{A+B} = \aleph_{A+B,A+B}, \\
&\aleph_{A,B+C} = \aleph_{A,B} + \aleph_{A,C}, \quad \aleph_{A,B-C} = \aleph_{A,B} - \aleph_{A,C}, \\
&\aleph_{A+(D+E),B+C} = \aleph_{A,B+C} + \aleph_{D+E,B+C}, \\
&\aleph_{A-D,B+C} = \aleph_{A,B+C} + \aleph_{-D,B+C} \\
&\aleph_{A-B,-(C-D)} + \aleph_{A-B,A-B} = \aleph_{A-B,(A-B)-(C-D)}, \quad \aleph_{A+B,-(C+D)} + \aleph_{A+B,A+B} = \aleph_{A+B,(A+B)-(C+D)} \\
&\aleph_{A-B,C-D} + \aleph_{A-B,A-B} = \aleph_{A-B,(A-B)+(C-D)}, \quad \aleph_{A+B,C+D} + \aleph_{A+B,A+B} = \aleph_{A+B,(A+B)+(C+D)}
\end{aligned} \qquad \text{(S59)}$$

**Supporting Information 6: J-DNP enhancements for other kinds of biradical/nuclear systems**

Table 1 in the main text focused on one combination of electron and nuclear spin coupling parameters, leading to the features noted in the paper. The two electrons had identical *g*-tensors, with electron and nuclear Zeeman couplings made anisotropic for the sake of realism. This section summarized four additional sets of combinations, as per the parameters summarized in Table S3. These include (i) the same coupling parameters as in Table 1 but for a different placement of the nucleus, which was now assumed devoid of chemical shift anisotropy; (ii) same parameters as in Table 1 but now with electron sites endowed with identical rhombic *g*-tensors; (iii) same parameters as in Table 1 but now with the electrons devoid of *g*-anisotropies; (iv) same parameters as in Table 1 but now with electron sites endowed with different isotropic *g*-tensors. The systems in the Table S3 thus contain axial, rhombic and isotropic *g*-tensors –in the latter case with coinciding and non-coinciding isotropic values. For the sake of conciseness, only the time-domain J-DNP transient enhancements were calculated for these scenarios –using the single optimal $J_{ex} = +(\omega_E + \omega_N)$ but as function of $B_o$ and $\tau_c$. These are shown in Figures S8-S11. Note how similar is the behaviour for all these systems, when compared with that shown in Figure 3.

**Table S3:** Biradical / proton magnetic resonance parameters used in the simulations shown in Figures S8-S11. Each electron in the biradical had its parameters modelled on a trityl center, and the nucleus was placed along the linker closer to one of the electrons. $B_o$, $J_{ex}$ and $\tau_c$ for the biradical/proton triad were set as described in the figures; all other hyperfine coupling parameters relied on the distances.

| Parameter | System i | System ii | System iii | System iv |
|---|---|---|---|---|
| $^1$H chemical shift tensor, ppm | [10  10  10] | [5  10  20] | [5  10  20] | [5  10  20] |
| *g*-tensor[1] for the electron 1 and 2, Bohr magneton | [2.0032 2.0032 2.0026] | [2.0030 2.0025 2.0020] | [2.0032 2.0032 2.0032] | $g_1$=[2.0032 2.0032 2.0032] $g_2$=[2.0027 2.0027 2.0027] |
| $^1$H coordinates, [x y z], Å | [-3 0.5 1.3] | [-3 0.5 1.3] | [-3 0.5 1.3] | [-3 0.5 1.3] |
| Electron 1 coordinates, [x y z], Å | [0  0  -9.37] | [0  0  -9.37] | [0  0  -9.37] | [0  0  -9.37] |
| Electron 2 coordinates, [x y z], Å | [0  0  9.37] | [0  0  9.37] | [0  0  9.37] | [0  0  9.37] |
| Scalar relaxation modulation depth /GHz | 3 | 3 | 3 | 1 |
| Scalar relaxation modulation time, ps | 1 | 1 | 1 | 1 |
| Temperature/ K | 298 | 298 | 298 | 298 |

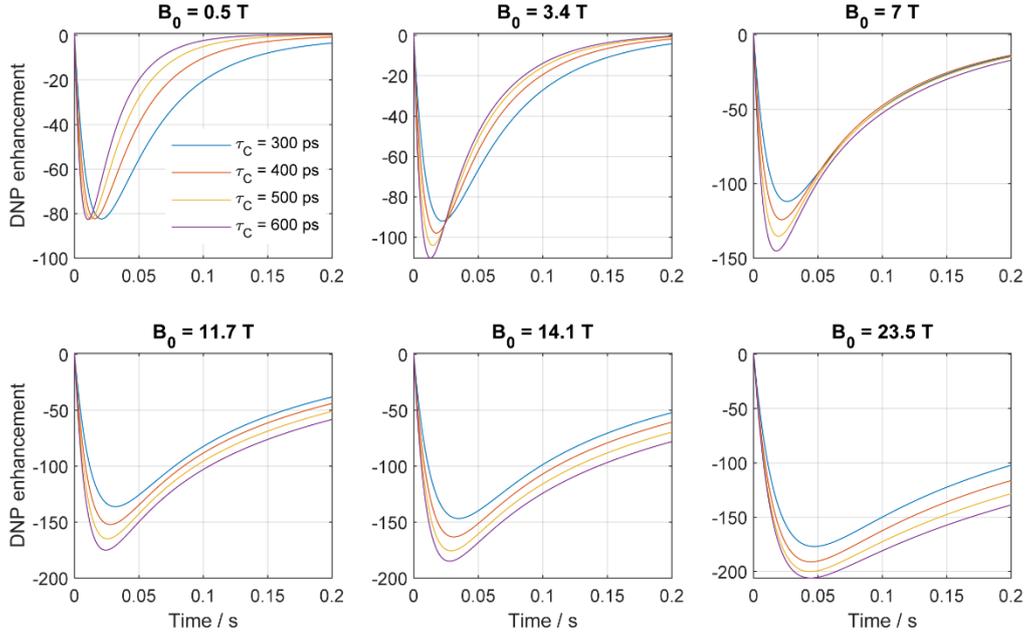

**Fig. S8:** Time domain simulations showing the evolution of the transient J-DNP enhancement as a function of $B_0$ and of $\tau_C$. For all fields $J_{ex}$ was tuned to $+(\omega_E + \omega_N)$, using the parameters of the system (i) in Table S3.

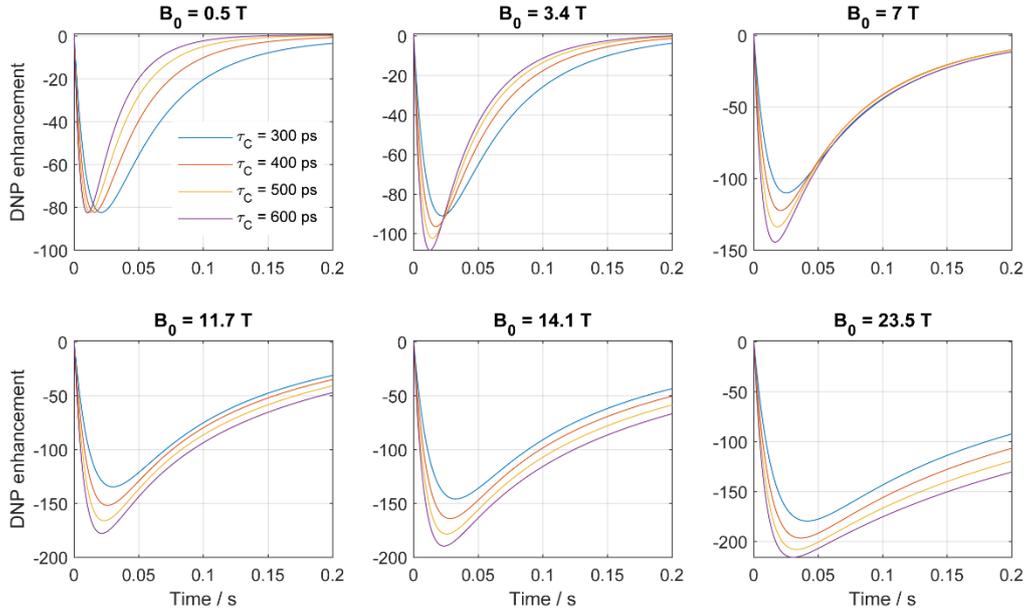

**Fig. S9:** Time domain simulations showing the evolution of the transient J-DNP enhancement as a function of $B_0$ and of $\tau_C$. For all fields $J_{ex}$ was tuned to $+(\omega_E + \omega_N)$, using the parameters of the system (ii) in Table S3.

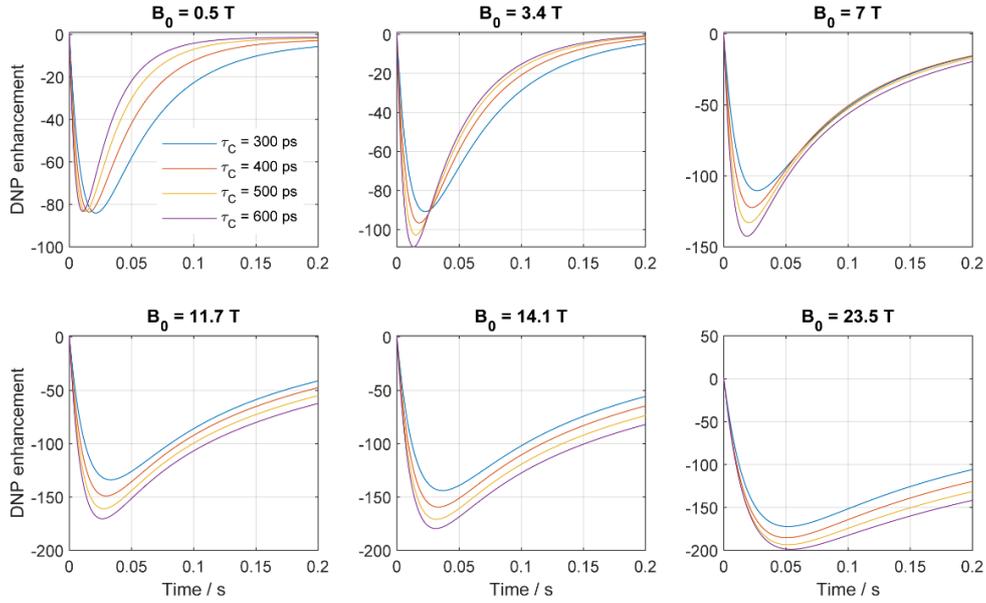

**Fig. S10:** Time domain simulations showing the evolution of the transient J-DNP enhancement as a function of $B_0$ and of $\tau_C$. For all fields $J_{ex}$ was tuned to $+(\omega_E + \omega_N)$, using the parameters of the system (iii) in Table S3.

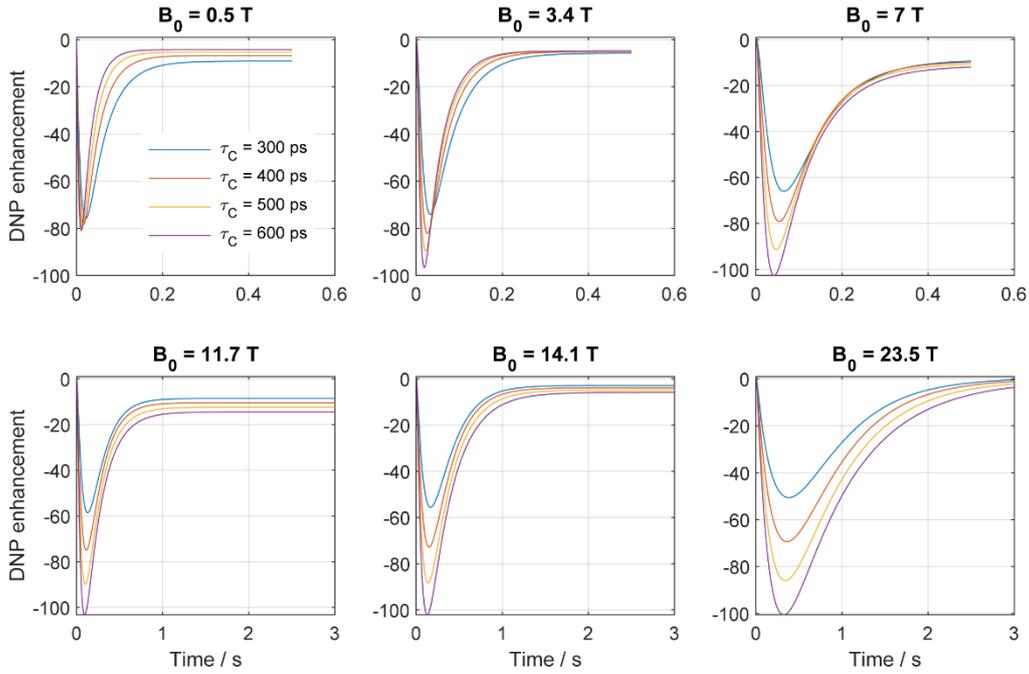

**Fig. S11:** Time domain simulations showing the evolution of the transient J-DNP enhancement as a function of $B_0$ and of $\tau_C$. For all fields $J_{ex}$ was tuned to $+(\omega_E + \omega_N)$, using the parameters of the system (iv) in Table S3.

Despite the noted similarity among all these cases, it is important to remark that cases will also arise where the J-DNP enhancement will be "killed", Figures S12 - S14 include three of such instances. In the first of these, the nucleus is symmetrically placed in-between two identical electrons, that would otherwise lead to enhancement; this makes the differential "CIDNP-like" effect stop working. J-DNP requires differential hyperfine couplings driving a differential relaxation-based "nuclear spin-state filter", in their absence, for instance if the nucleus is symmetrically placed between the two electrons (Supporting Figure S12), no enhancement results. The second case involves two electron sites endowed with different anisotropic g-tensors (Supporting Figure S13); in such instance the $\aleph_{\Delta G, \Delta G \pm \Delta HF}$ terms overtake the $\Delta^2_{\text{HFC}}$ in Eqs. (5) – (12), robbing J-DNP for its efficiency even when $J_{\text{ex}} = \pm(\omega_E + \omega_N)$. Notice that this does not happen when $g_{1,\text{iso}} = g_{2,\text{iso}}$ (Supporting Figure S11), as the $\aleph_{\Delta G, \Delta G \pm \Delta HF}$ terms still remain then smaller than $\Delta^2_{\text{HFC}}$. Finally, the enhancement will tend to zero if the nucleus remains too distant from the biradical: for instance, a proton placed 20 Å away from the biradical that may require over 1 s to achieve significant polarization gains, a time by which the DNP effect will lose against competing pathways. Interestingly, despite J-DNP's origin in effects related to second-rank spherical anisotropies, its nuclear enhancement never changes sign (Supporting Figure S14).

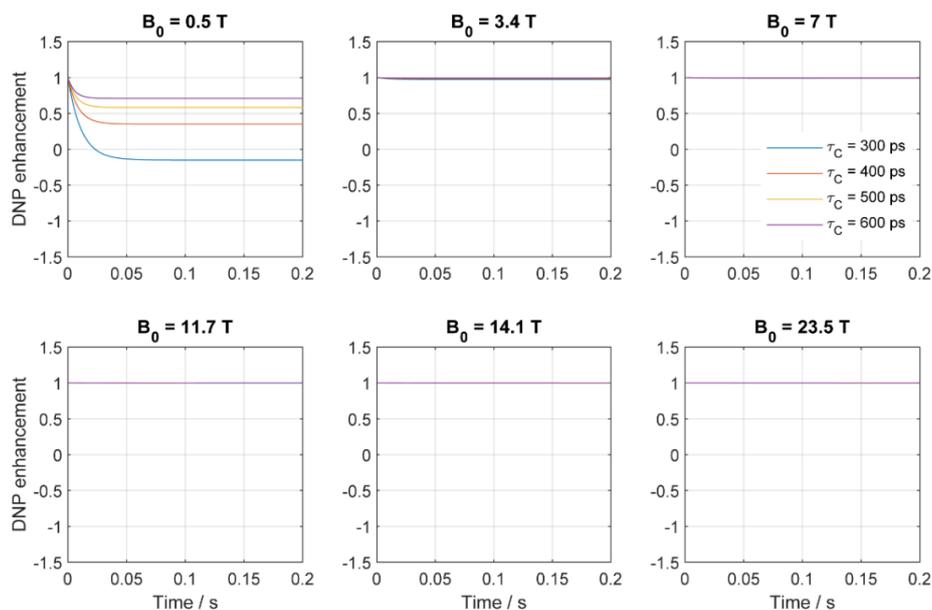

**Fig. S12:** Time domain simulations showing the evolution of the transient J-DNP enhancement as a function of $B_0$ and of $\tau_C$. For all fields $J_{\text{ex}}$ was tuned to $+(\omega_E + \omega_N)$, using the parameters of the system in Table 1, but with the proton placed symmetrically in the biradical's centre (i.e, at [0 0 0] Å).

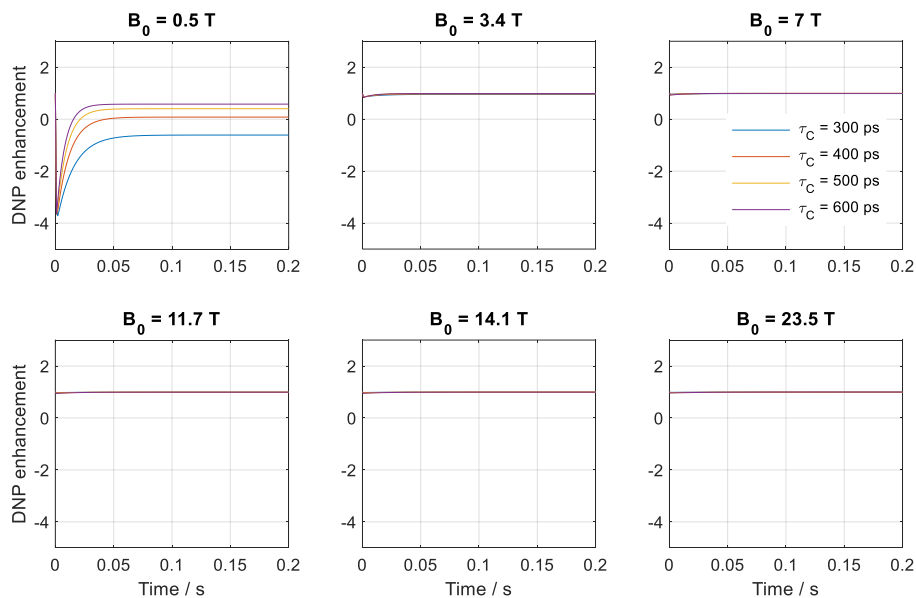

**Fig. S13:** Time domain simulations showing the evolution of the transient J-DNP enhancement as a function of $B_0$ and of $\tau_C$. For all fields $J_{\text{ex}}$ was tuned to $+(\omega_E + \omega_N)$, using the same parameters as in Table 1 but now with electron sites endowed with different anisotropic g-tensors equal to: $g_1$=[2.0032 2.0032 2.0026] and $g_2$=[2.0032 2.0032 2.0023].

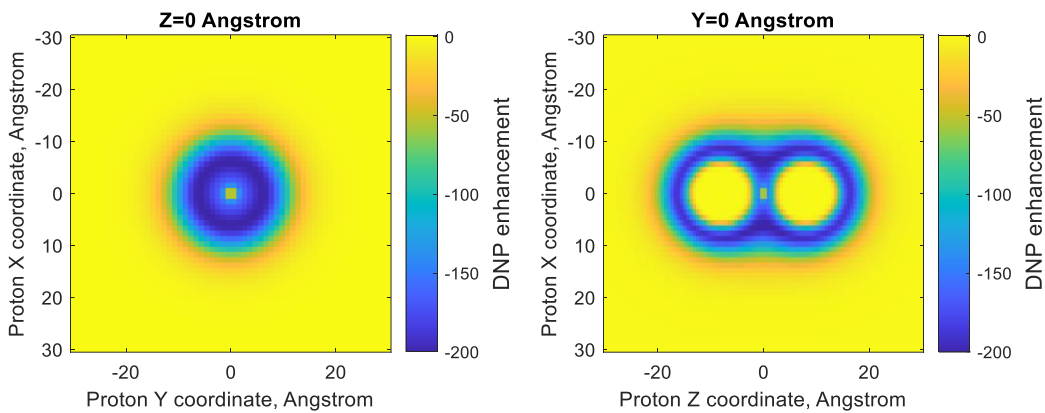

**Fig. S14:** Maximum enhancement (amplitude of $\hat{N}_Z$ normalized to the thermal equilibrium value of a single proton at the same magnetic field) achieved within 20 ms of microwave irradiation at the electron Larmor frequency, as a function of random $^1$H-coordinates surrounding a model biradical. Parameters included $B_0$ = 14.08 T, $\tau_c$ = 500 ps of the biradical/proton triad, other conditions as given in Table 1. Notice the negative enhancement displayed by all positions surrounding the radical. This is important, as otherwise the spatial averaging brought about by molecular translations, could end up being smaller or even zero.

## References

1   Owenius, R., Eaton, G. R. & Eaton, S. S. Frequency (250 MHz to 9.2 GHz) and viscosity dependence of electron spin relaxation of triarylmethyl radicals at room temperature. *J Magn Reson* **172**, 168, (2005).

2   Yong, L. *et al.* Electron spin relaxation of triarylmethyl radicals in fluid solution. *J Magn Reson* **152**, 156, (2001).

3   Blicharski, J. Nuclear magnetic relaxation by anisotropy of the chemical shift. *Z Naturforsch Pt A* **27**, 1456 (1972).

4   Kuprov, I., Wagner-Rundell, N. & Hore, P. Bloch-Redfield-Wangsness theory engine implementation using symbolic processing software. *J Magn Reson* **184**, 196 (2007).

5   Ardenkjaer-Larsen, J. H. *et al.* EPR and DNP properties of certain novel single electron contrast agents intended for oximetric imaging. *J Magn Reson* **133**, 1, (1998).

6   Hausser, K. H. & Stehlik, D. Dynamic nuclear polarization in liquids, in Advances in Magnetic and Optical Resonance. Vol. 3, p.79-139 (Elsevier, 1968).

7   Griesinger, C. *et al.* Dynamic nuclear polarization at high magnetic fields in liquids. *Prog Nucl Magn Reson Spectrosc* **64**, 4, (2012).

8   Prisner, T., Denysenkov, V. & Sezer, D. Liquid state DNP at high magnetic fields: Instrumentation, experimental results and atomistic modelling by molecular dynamics simulations. *J Magn Reson* **264**, 68, (2016).

9   Parigi, G., Ravera, E., Bennati, M. & Luchinat, C. Understanding Overhauser Dynamic Nuclear Polarisation through NMR relaxometry. *Mol Phys* **117**, 888, (2019).

10  Levien, M., Hiller, M., Tkach, I., Bennati, M. & Orlando, T. Nitroxide Derivatives for Dynamic Nuclear Polarization in Liquids: The Role of Rotational Diffusion. *J Phys Chem Lett* **11**, 1629, (2020).

11  Wind, R. A. & Ardenkjaer-Larsen, J. H. 1H DNP at 1.4 T of water doped with a triarylmethyl-based radical. *J Magn Reson* **141**, 347-354, (1999).

12  Hofer, P. *et al.* Field dependent dynamic nuclear polarization with radicals in aqueous solution. *J Am Chem Soc* **130**, 3254, (2008).

13  Hu, K. N., Debelouchina, G. T., Smith, A. A. & Griffin, R. G. Quantum mechanical theory of dynamic nuclear polarization in solid dielectrics. *J Chem Phys* **134**, 125105, (2011).

14   Bajaj, V. S. *et al.* 250 GHz CW gyrotron oscillator for dynamic nuclear polarization in biological solid state NMR. *J Magn Reson* **189**, 251, (2007).
15   Vega, S. Fictitious spin 1/2 operator formalism for multiple quantum NMR. *J Chem Phys* **68**, 5518 (1978).
16   Ernst, R. R., Bodenhausen, G. & Wokaun, A. Principles of Nuclear Magnetic Resonance in One and Two Dimensions.  p.610 (Oxford: Clarendon Press, 1987).
17   Pileio, G. Singlet NMR methodology in two-spin-1/2 systems. *Prog Nucl Magn Reson Spectrosc* **98-99**, 1, (2017).
18   Bengs, C. & Levitt, M. H. SpinDynamica: Symbolic and numerical magnetic resonance in a Mathematica environment. *Magn Reson Chem* **56**, 374, (2018).
19   Redfield, A. *The theory of relaxation processes, in Advances in Magnetic and Optical Resonance*. Vol. 1, p.1-32 (Elsevier, 1965).
20   Goldman, M. Formal theory of spin--lattice relaxation. *J Magn Reson* **149**, 160, (2001).